\documentclass[twocolumn]{emulateapj}

\usepackage{amsmath,amssymb}
\usepackage{graphicx}
\usepackage{subfigure}
\usepackage{color}
\usepackage{ulem}
\usepackage{booktabs}

\newcommand{\Msun}{{\rm M_\odot}}

\newcommand{\diff}{{\rm {d}}}
\newcommand{\dd}{{\; \rm {d}}}
\newcommand{\lpt}{\textsc{2lpt}}
\newcommand{\za}{\textsc{za}}

\newcommand{\gadgettwo}{\textsc{Gadget-2}}
\newcommand{\rockstar}{\textsc{Rockstar}}

\newcommand{\nbody}{{\textit{N}-body}}

\renewcommand{\digamma}{{\psi_{0}}}
\shorttitle{Halo Populations in \lpt\ and \za\ Simulations}
\shortauthors{Sissom, Holley-Bockelmann, Sinha}

\begin{document}

\title{Early Growth in a Perturbed Universe:\\  Exploring Dark Matter Halo Populations in \lpt\ and \za\ Simulations}
\author{Daniel J. Sissom\altaffilmark{1,2}, Kelly Holley-Bockelmann\altaffilmark{1,3}, Manodeep Sinha\altaffilmark{1,4}}

\altaffiltext{1}{Dept.\ of Physics and Astronomy, Vanderbilt University, Nashville, TN, 37235}
\altaffiltext{2}{Email:  {\tt daniel.j.sissom@vanderbilt.edu}}
\altaffiltext{3}{Email:  {\tt k.holley@vanderbilt.edu}}
\altaffiltext{4}{Email:  {\tt manodeep.sinha@vanderbilt.edu}}

\begin{abstract}
	We study the structure and evolution of dark matter halos from $z = 300$ to $z = 6$ for two cosmological \nbody\ simulation initialization techniques.  While the second order Lagrangian perturbation theory (\lpt) and the Zel'dovich approximation (\za) both produce accurate present day halo mass functions, earlier collapse of dense regions in \lpt\ can result in larger mass halos at high redshift.  We explore the differences in dark matter halo mass and concentration due to initialization method through three \lpt\ and three \za\ initialized cosmological simulations.  We find that \lpt\ induces more rapid halo growth, resulting in more massive halos compared to \za.  This effect is most pronounced for high mass halos and at high redshift, with a fit to the mean normalized difference between \lpt\ and \za\ halos as a function of redshift of $\mu_{\Delta M_{\mathrm{vir}}} = (7.88 \pm 0.17) \times 10^{-3} z - (3.07 \pm 0.14) \times 10^{-2}$.  Halo concentration is, on average, largely similar between \lpt\ and \za, but retains differences when viewed as a function of halo mass.  For both mass and concentration, the difference between typical individual halos can be very large, highlighting the shortcomings of \za-initialized simulations for high-$z$ halo population studies.
\end{abstract}

\maketitle

%%%%%%%%%%%%%%%%%%%%%%%%%%%%%%%%%%%%%%%%%%%%%%%%%%%%%%%%%%%%%%%%%%%%%%%%%%%%%%%%
%
% Introduction
%
%%%%%%%%%%%%%%%%%%%%%%%%%%%%%%%%%%%%%%%%%%%%%%%%%%%%%%%%%%%%%%%%%%%%%%%%%%%%%%%%

\section{Introduction}
\label{sec:introduction}

%%%%%%%%%%%%%%%%%%%%%%%%%%%%%%%%%%%%%%%%%%%%%%%%%%%%%%%%%%%%%%%%%%%%%%%%%%%%%%%%

%~~~~~~~~~~~~~~~~~~~~~~~~~~~~~~~~~~~~~~~~~~~~~~~~~~~~~~~~~~~~~~~~~~~~~~~~~~~~~~~
% Structure formation in the pre-reionization epoch
%~~~~~~~~~~~~~~~~~~~~~~~~~~~~~~~~~~~~~~~~~~~~~~~~~~~~~~~~~~~~~~~~~~~~~~~~~~~~~~~

The pre-reionization epoch is a time of significant evolution of early structure in the Universe.  Rare density peaks in the otherwise smooth dark matter (DM) sea lead to the collapse and formation of the first dark matter halos.  For example, at $z = 20$, $10^{7}~\mathrm{M}_{\odot}$ halos are $\sim 4\sigma$ peaks, and $10^{8}~\mathrm{M}_{\odot}$ halos, candidates for hosting the first supermassive black hole seeds, are $\sim 5\sigma$ peaks.

These early-forming dark matter halos provide an incubator for the baryonic processes that allow galaxies to form and transform the surrounding IGM.  Initial gas accretion can lead to the formation of the first Pop-III stars \citep{1986MNRAS.221...53C, 1997ApJ...474....1T, 2000ApJ...540...39A, 2002Sci...295...93A}, which, upon their death, can collapse into the seeds for supermassive black holes (SMBHs) \citep{2001ApJ...551L..27M, 2003MNRAS.340..647I, 2009ApJ...701L.133A, 2012ApJ...754...34J} or enrich the surrounding medium with metals through supernovae \citep{2002ApJ...567..532H, 2003ApJ...591..288H}.  The radiation from early quasars \citep{1987ApJ...321L.107S, 1999ApJ...514..648M, 2001AJ....122.2833F}, Pop-III stars \citep{1997ApJ...486..581G, 2003ApJ...584..621V, 2006ApJ...639..621A}, and proto-galactic stellar populations \citep{2012ApJ...752L...5B, 2012MNRAS.423..862K} all play a key role in contributing to re-ionizing the Universe by around $z = 6$ \citep{2001PhR...349..125B}.  Additionally, halo mergers can drastically increase the temperature of halo gas through shock heating, increasing X-ray luminosity \citep{2009MNRAS.397..190S} and unbinding gas to form the warm-hot intergalactic medium \citep{2008SSRv..134..141B, 2010MNRAS.405L..31S, 2012MNRAS.425.2974T}.

%~~~~~~~~~~~~~~~~~~~~~~~~~~~~~~~~~~~~~~~~~~~~~~~~~~~~~~~~~~~~~~~~~~~~~~~~~~~~~~~
% Current knowledge of high z mass, concentration, and density
%~~~~~~~~~~~~~~~~~~~~~~~~~~~~~~~~~~~~~~~~~~~~~~~~~~~~~~~~~~~~~~~~~~~~~~~~~~~~~~~

Since the pre-reionization era is such a critical epoch in galaxy evolution, much effort is expended to characterize the dark matter distribution accurately.  Statistical measures of the DM halo population, such as the halo mass function, are employed to take a census of the collapsed halos, while 3-point correlation functions are used to describe the clustering of these halos as a probe of cosmology.  Detailed analysis of the structure of individual halos involves characterizing the DM halo mass and density profile.

There are a number of ways to define a halo's mass, the subtleties of which become significant for mass-sensitive studies, such as the halo mass function \citep{1974ApJ...187..425P, 2007MNRAS.374....2R, 2006ApJ...642L..85H, 2007ApJ...671.1160L}.  For a review, see, e.g., \citet{2001AA...367...27W} and references therein.  Additionally, see \citet{2005RvMP...77..207V} and references therein for a more observationally-focused discussion.  From a simulation standpoint, however, the two most common ways to obtain halo mass are through either spherical overdensity or friends-of-friends (FOF) techniques.  The spherical overdensity method identifies regions above a certain density threshold, either with respect to the critical density $\rho_{c} = 3 H^{2} / 8 \pi G$ or the background density $\rho_{b} = \Omega_{m} \rho_{c}$, where $\Omega_{m}$ is the matter density of the universe.  The mass is then the mass enclosed in a sphere of some radius with mean density $\Delta \rho_{c}$, where $\Delta$ commonly ranges from $\sim 100$ to $\sim 500$.  Alternatively, the FOF method finds particle neighbors and neighbors of neighbors defined to be within some separation distance \citep{1984MNRAS.206..529E, 1985ApJ...292..371D}.  Halo mass, then, is simply the sum of the masses of the linked particles.

The density profile of a DM halo is most often modeled with the NFW \citep{1996ApJ...462..563N} profile:
\begin{equation} \label{eq:nfw_profile}
	\rho(r) = \frac{ \rho_{0} }{ \frac{ r }{ R_{s}} \left( 1 + \frac{r}{R_{s}} \right)^{2} },
\end{equation}
where $\rho_{0}$ is the characteristic density, and the scale radius $R_{s}$ is the break radius between the inner $\sim r^{-1}$ and outer $\sim r^{-3}$ density profiles.  The NFW density profile is quantified by the halo concentration $c \equiv R_{\mathrm{vir}} / R_{s}$.  $R_{\mathrm{vir}}$ is the halo virial radius, which is often defined as the radius at which the average interior density is some factor $\Delta_{c}$ times the critical density of the universe $\rho_{c}$, where $\Delta_{c}$ is typically $\sim 200$.  Concentration may also be obtained for halos modeled with the Einasto \citep{1989AA...223...89E} profile.  However, while halo profiles can be better approximated by the Einasto profile \citep{2004MNRAS.349.1039N, 2010MNRAS.402...21N, 2008MNRAS.387..536G}, the resulting concentrations display large fluctuations due to the smaller curvature of the density profile around the scale radius \citep{2012MNRAS.423.3018P}.

Generally, at low redshift, low mass halos are more dense than high mass halos \citep{1997ApJ...490..493N}, and concentration decreases with redshift and increases in dense environments \citep{2001MNRAS.321..559B}.  \citet{2007MNRAS.381.1450N} additionally find that concentration decreases with halo mass.  Various additional studies have explored concentration's dependence on characteristics of the power spectrum \citep{2001ApJ...554..114E}, cosmological model \citep{2008MNRAS.391.1940M}, redshift \citep{2008MNRAS.387..536G, 2011MNRAS.411..584M}, and halo merger and mass accretion histories \citep{2002ApJ...568...52W, 2003MNRAS.339...12Z, 2009ApJ...707..354Z}.  For halos at high redshift, \citet{2011ApJ...740..102K} find that concentration reverses and increases with mass for high mass halos, while \citet{2012MNRAS.423.3018P} additionally find that concentration's dependence on mass and redshift is better correlated with $\sigma(M,z)$, the RMS fluctuation amplitude of the linear density field.

%~~~~~~~~~~~~~~~~~~~~~~~~~~~~~~~~~~~~~~~~~~~~~~~~~~~~~~~~~~~~~~~~~~~~~~~~~~~~~~~
% Simulation initialization
%~~~~~~~~~~~~~~~~~~~~~~~~~~~~~~~~~~~~~~~~~~~~~~~~~~~~~~~~~~~~~~~~~~~~~~~~~~~~~~~

Cosmological simulations that follow the initial collapse of dark matter density peaks into virialized halos often neglect to consider the nuances of initialization method.  Despite much effort in characterizing the resulting DM structure, comparatively less attention is paid to quantifying the effect of the initialization and simulation technique used to obtain the DM distribution.  The subtle $\mathcal{O}(10^{-5})$ density perturbations in place at the CMB epoch are vulnerable to numerical noise and intractable to simulate directly.  Instead, a displacement field is applied to the particles to evolve them semi-analytically, nudging them from their initial positions to an approximation of where they should be at a more reasonable starting redshift for the numerical simulation.  Starting at a later redshift saves computation time as well as avoiding interpolation systematics and round-off errors \citep{2007ApJ...671.1160L}.

%~~~~~~~~~~~~~~~~~~~~~~~~~~~~~~~~~~~~~~~~~~~~~~~~~~~~~~~~~~~~~~~~~~~~~~~~~~~~~~~
% 2LPT and ZA
%~~~~~~~~~~~~~~~~~~~~~~~~~~~~~~~~~~~~~~~~~~~~~~~~~~~~~~~~~~~~~~~~~~~~~~~~~~~~~~~

The two canonical frameworks for the initial particle displacement involved in generating simulation initial conditions are the Zel'dovich approximation \citep[\za,][]{1970AA.....5...84Z} and 2nd-order Lagrangian Perturbation Theory \citep[\lpt,][]{1994MNRAS.267..811B, 1994AA...288..349B, 1995AA...296..575B, 1998MNRAS.299.1097S}.  \za\ initial conditions displace initial particle positions and velocities via a linear field \citep{1983MNRAS.204..891K, 1985ApJS...57..241E}, while \lpt\ initial conditions add a second-order correction term to the expansion of the displacement field \citep{1998MNRAS.299.1097S, 2005ApJ...634..728S, 2010MNRAS.403.1859J}.

Following \citet{2010MNRAS.403.1859J}, we briefly outline \lpt\ and compare it to \za.  In \lpt, a displacement field $\boldsymbol{\Psi}(\boldsymbol{q})$ is applied to the initial positions $\boldsymbol{q}$ to yield the Eulerian final comoving positions
\begin{equation} \label{eq:displacement}
	\boldsymbol{x} = \boldsymbol{q} + \boldsymbol{\Psi}.
\end{equation}
The displacement field is given in terms of two potentials $\phi^{(1)}$ and $\phi^{(2)}$:
\begin{equation} \label{eq:potentials}
	\boldsymbol{x} = \boldsymbol{q} - D_{1} \boldsymbol{\nabla}_{q} \phi^{(1)} + D_{2} \boldsymbol{\nabla}_{q} \phi^{(2)},
\end{equation}
with linear growth factor $D_{1}$ and second-order growth factor $D_{2} \approx -3 D_{1}^{2} / 7$.  The subscripts $q$ refer to partial derivatives with respect to the Lagrangian coordinates $\boldsymbol{q}$.  Likewise, the comoving velocities are given, to second order, by
\begin{equation} \label{eq:velocity}
	\boldsymbol{v} =  - D_{1} f_{1} H \boldsymbol{\nabla}_{q} \phi^{(1)} + D_{2} f_{2} H \boldsymbol{\nabla}_{q} \phi^{(2)},
\end{equation}
with Hubble constant $H$ and $f_{i} = \diff\, \mathrm{ln}\, D_{i} / \diff\, \mathrm{ln}\, a$, where $a$ is the expansion factor.  The relations $f_{1} \approx \Omega_{m}^{5/9}$ and $f_{2} \approx 2 \Omega_{m}^{6/11}$, with matter density $\Omega_{m}$, apply for flat models with a non-zero cosmological constant \citep{1995AA...296..575B}.  The $f_{1}$, $f_{2}$, and $D_{2}$ approximations here are very accurate for most actual $\Lambda$CDM initial conditions, as $\Omega_{m}$ is close to unity at high starting redshift \citep{2010MNRAS.403.1859J}.  We may derive $\phi^{(1)}$ and $\phi^{(2)}$ by solving a pair of Poisson equations:
\begin{equation} \label{eq:poisson1}
	\nabla_{q}^{(1)}(\boldsymbol{q}) = \delta^{(1)}(\boldsymbol{q}),
\end{equation}
with linear overdensity $\delta^{(1)}(\boldsymbol{q})$, and
\begin{equation} \label{eq:poisson2}
	\nabla_{q}^{(2)}(\boldsymbol{q}) = \delta^{(2)}(\boldsymbol{q}).
\end{equation}
The second order overdensity $\delta^{(2)}(\boldsymbol{q}$) is related to the linear overdensity field by
\begin{equation} \label{eq:second-order_overdensity}
	\delta^{(2)}(\boldsymbol{q}) = \sum_{i > j} \left\{ \phi_{,ii}^{(1)}(\boldsymbol{q}) \phi_{,jj}^{(1)}(\boldsymbol{q}) - \left[ \phi_{,ij}^{(1)}(\boldsymbol{q}) \right]^{2} \right\},
\end{equation}
where $\phi_{,ij} \equiv \partial^{2} \phi / \partial q_{i} \partial q_{j}$.  For initial conditions from \za, or first-order Lagrangian initial conditions, the $\phi^{(2)}$ terms of Equations~\ref{eq:potentials} and \ref{eq:velocity} are ignored.

%~~~~~~~~~~~~~~~~~~~~~~~~~~~~~~~~~~~~~~~~~~~~~~~~~~~~~~~~~~~~~~~~~~~~~~~~~~~~~~~
% Transients
%~~~~~~~~~~~~~~~~~~~~~~~~~~~~~~~~~~~~~~~~~~~~~~~~~~~~~~~~~~~~~~~~~~~~~~~~~~~~~~~

In theory, non-linear decaying modes, or transients, will be damped as $1 / a$ in \za.  In \lpt, however, transients are damped more quickly as $1 / a^{2}$.  It should be expected, then, that structure in \lpt\ will be accurate after fewer $e$-folding times than in \za\ \citep{1998MNRAS.299.1097S, 2006MNRAS.373..369C, 2010MNRAS.403.1859J}.  The practical result is that high-$\sigma$ DM density peaks at high redshift are suppressed in \za\ compared with \lpt\ for a given starting redshift \citep{2006MNRAS.373..369C}.  While differences in ensemble halo properties, such as the halo mass function, between simulation initialization methods are mostly washed away by $z=0$ \citep{1998MNRAS.299.1097S}, trends at earlier redshifts are less studied \citep{2007ApJ...671.1160L}.

%~~~~~~~~~~~~~~~~~~~~~~~~~~~~~~~~~~~~~~~~~~~~~~~~~~~~~~~~~~~~~~~~~~~~~~~~~~~~~~~
% In this paper...
%~~~~~~~~~~~~~~~~~~~~~~~~~~~~~~~~~~~~~~~~~~~~~~~~~~~~~~~~~~~~~~~~~~~~~~~~~~~~~~~

In this paper, we explore the effects of \za\ and \lpt\ on the evolution of halo populations at high redshift.  It is thought that \lpt\ allows initial DM overdensities to get a ``head start'' compared with \za, allowing earlier structure formation, more rapid evolution, and larger possible high-mass halos for a given redshift.  We explore this possibility by evolving a suite of simulations from $z = 300$ to $z = 6$ and comparing the resulting differences in halo properties arising from initialization with \za\ and \lpt\ in these these otherwise identical simulations.

We discuss the simulations, halo finding, and analysis methods in Section~\ref{sec:methods}, results in Section~\ref{sec:results}, implications, caveats, and future work in Section~\ref{sec:discussion}, and a summary of our results and conclusions in Section~\ref{sec:conclusion}.

%%%%%%%%%%%%%%%%%%%%%%%%%%%%%%%%%%%%%%%%%%%%%%%%%%%%%%%%%%%%%%%%%%%%%%%%%%%%%%%%
%
% Numerical Methods
%
%%%%%%%%%%%%%%%%%%%%%%%%%%%%%%%%%%%%%%%%%%%%%%%%%%%%%%%%%%%%%%%%%%%%%%%%%%%%%%%%

\section{Numercial Methods}
\label{sec:methods}

%%%%%%%%%%%%%%%%%%%%%%%%%%%%%%%%%%%%%%%%%%%%%%%%%%%%%%%%%%%%%%%%%%%%%%%%%%%%%%%%

%~~~~~~~~~~~~~~~~~~~~~~~~~~~~~~~~~~~~~~~~~~~~~~~~~~~~~~~~~~~~~~~~~~~~~~~~~~~~~~~
% Simulations
%~~~~~~~~~~~~~~~~~~~~~~~~~~~~~~~~~~~~~~~~~~~~~~~~~~~~~~~~~~~~~~~~~~~~~~~~~~~~~~~

We use the \nbody\ tree/SPH code \gadgettwo\ \citep{2001NewA....6...79S, 2005MNRAS.364.1105S} to evolve six dark matter--only cosmological volumes from $z_{start} = 300$ to $z = 6$ in a $\rm \Lambda CDM$ universe.  Each simulation is initialized using WMAP-5 \citep{2009ApJS..180..330K} parameters.  For each of the three simulation pairs, we directly compare \lpt\ and \za\ by identically sampling the CMB transfer function and displacing the initial particle positions to the same starting redshift using \lpt\ and \za.  The three sets of simulations differ only by the initial phase sampling random seed.  Each volume contains $512^{3}$ particles in a 10 $h^{-1}$ Mpc box.  Following \citet{2010ApJ...715..104H}, we choose conservative simulation parameters in order to ensure high accuracy in integrating the particle positions and velocities.  We have force accuracy of 0.002, integration accuracy of 0.00125, and softening of $0.5\ h^{-1}\ \mathrm{kpc}$, or $1 / 40$ of the initial mean particle separation.  We use a uniform particle mass of $5.3 \times 10^{5} h^{-1} \Msun$.  Full simulation details are discussed in \citet{2012ApJ...761L...8H}.

One facet often overlooked when setting up an \nbody\ simulation is an appropriate starting redshift, determined by box size and resolution \citep{2007ApJ...671.1160L}.  As \lpt\ more accurately displaces initial particle positions and velocities, initialization with \lpt\ allows for a later starting redshift compared with an equivalent \za-initialized simulation.  However, many \za\ simulations do not take this into account, starting from too late an initial redshift and not allowing enough e-foldings to adequately dampen away numerical transients. \citep{2006MNRAS.373..369C, 2010MNRAS.403.1859J}.  In order to characterize an appropriate starting redshift, the relation between the initial RMS particle displacement and mean particle separation must be considered.  The initial RMS displacement $\Delta_{\mathrm{rms}}$ is given by
\begin{equation}
	\Delta_{\mathrm{rms}}^{2} = \frac{4 \pi}{3} \int_{k_{f}}^{k_{\mathrm{Ny}}} P(k, z_{\mathrm{start}}) \dd k,
\end{equation}
where $k_{f} = 2 \pi / L_{\mathrm{box}}$ is the fundamental mode, $L_{\mathrm{box}}$ is the simulation box size, $k_{\mathrm{Ny}} = \frac{1}{2} N k_{f}$ is the Nyquist frequency of an $N^{3}$ simulation, and $P(k, z_{\mathrm{start}})$ is the power spectrum at starting redshift $z_{\mathrm{start}}$.  In order to avoid the ``orbit crossings'' that reduce the accuracy of the initial conditions, $\Delta_{\mathrm{rms}}$ must be some factor smaller than the mean particle separation $\Delta_{p} = L_{\mathrm{box}} / N$ \citep{2012ApJ...761L...8H}.  For example, making orbit crossing a $\sim 10 \sigma$ event imposes $\Delta_{\mathrm{rms}} / \Delta_{p} = 0.1$.  However, for small-volume, high-resolution simulations, this quickly leads to impractical starting redshifts.  Continuing our example, satisfying $\Delta_{\mathrm{rms}} / \Delta_{p} \sim 0.1$ for a $10 h^{-1}$ Mpc, $512^{3}$ simulation suggests $z_{\mathrm{start}} \approx 799$.  Unfortunately, starting at such a high redshift places such a simulation well into the regime of introducing errors from numerical noise caused by roundoff errors dominating the smooth potential.  A more relaxed requirement of $\Delta_{\mathrm{rms}} / \Delta_{p} = 0.25$, which makes orbit crossing a $\sim 4\sigma$ event, yields $z_{\mathrm{start}} = 300$, which we adopt for this work.  For our small volume, the fundamental mode becomes non-linear at $z \sim 5$, after which, simulation results would become unreliable.  We therefore end our simulations at $z = 6$.

%~~~~~~~~~~~~~~~~~~~~~~~~~~~~~~~~~~~~~~~~~~~~~~~~~~~~~~~~~~~~~~~~~~~~~~~~~~~~~~~
% Rockstar
%~~~~~~~~~~~~~~~~~~~~~~~~~~~~~~~~~~~~~~~~~~~~~~~~~~~~~~~~~~~~~~~~~~~~~~~~~~~~~~~

For each of our six simulations, we use the 6-D phase space halo finder code \rockstar\ \citep{2013ApJ...762..109B} to identify spherical overdensity halos at each timestep.  \rockstar\ follows an adaptive hierarchical refinement of friends-of-friends halos in 6-D phase space, allowing determination of halo properties such as halo mass, position, virial radius, internal energy, and number of subhalos.  \rockstar\ tracks halos down to a threshold of around 20 particles, but we use a more conservative 100 particle threshold for our analysis.  We use all particles found within the virial radius to define our halos and their properties.

%~~~~~~~~~~~~~~~~~~~~~~~~~~~~~~~~~~~~~~~~~~~~~~~~~~~~~~~~~~~~~~~~~~~~~~~~~~~~~~~
% CrossMatch
%~~~~~~~~~~~~~~~~~~~~~~~~~~~~~~~~~~~~~~~~~~~~~~~~~~~~~~~~~~~~~~~~~~~~~~~~~~~~~~~

We identify companion halos between \lpt\ and \za\ simulations based on the highest fraction of matching particles contained in each at any given timestep.  We remove halo pairs where either one or both halos are considered subhalos (i.e. a halo must not be contained within another halo) and pairs with fewer than 100 particles in either \lpt\ or \za.  We are left with approximately 60,000 total halo pairs for our three boxes at $z = 6$.  With halo catalogs matched between simulations, we can compare properties of individual corresponding halos.  To mitigate the effects of cosmic variance on our small volumes, we ``stack'' the three simulation boxes for each initialization method, and combine the halos from each into one larger sample for our analysis.

%~~~~~~~~~~~~~~~~~~~~~~~~~~~~~~~~~~~~~~~~~~~~~~~~~~~~~~~~~~~~~~~~~~~~~~~~~~~~~~~
% Density Profiles
%~~~~~~~~~~~~~~~~~~~~~~~~~~~~~~~~~~~~~~~~~~~~~~~~~~~~~~~~~~~~~~~~~~~~~~~~~~~~~~~

\begin{figure}[t]
    \centering
    \includegraphics[width=\linewidth]{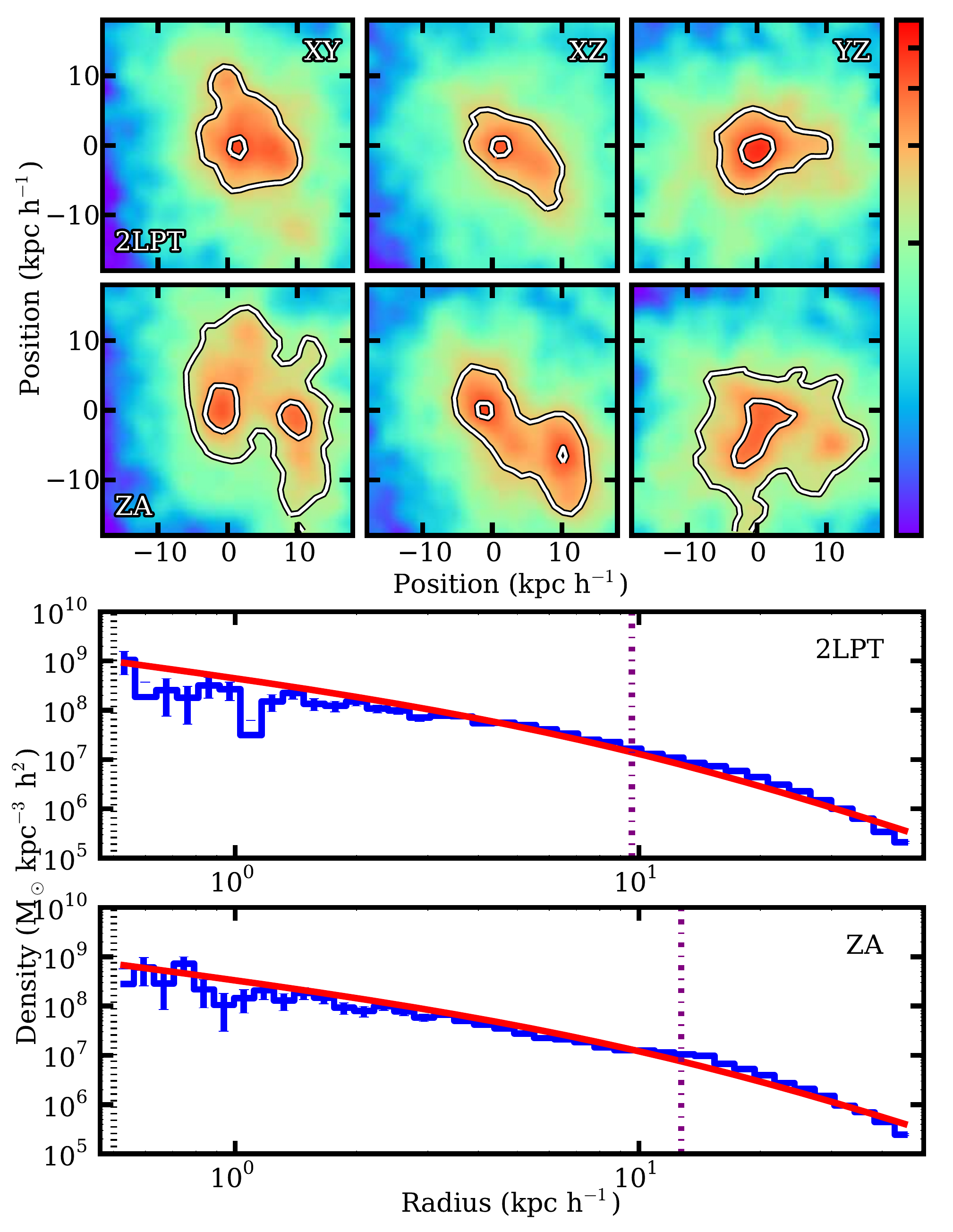}
	\caption[Comparison of matched \lpt\ and \za\ halos]{\footnotesize \textit{Top two rows:}  Density projections for two matching halos at $z = 6$.  The first and second row are \lpt\ and \za, respectively.  The halos appear to be either undergoing or have recently undergone a major merger.  The \lpt\ halo appears to be more relaxed and further along in the merger process, while the \za\ halo lags behind, still displaying two distinct cores.  The halos have masses of $5.95 \times 10^{9} \textrm{M}_{\odot}$ for \lpt\ and $5.85 \times 10^{9} \textrm{M}_{\odot}$ for \za.  \textit{Bottom two rows:}  Density profiles for the same two halos as above.  NFW profiles are fit to logarithmic radial bins of particle position and are overplotted as red curves.  The purple dot--dash lines mark the scale radii.  The black dotted lines mark the resolution limit of the simulations.}
    \label{fig:halo-pair}
\end{figure}

Halo concentration is derived from \rockstar's output for $R_{s}$ and $R_{\mathrm{vir}}$.  Here, $R_{\mathrm{vir}}$ is the virial radius as defined by \citet{1998ApJ...495...80B}.  Figure~\ref{fig:halo-pair} makes evident the difficulty in fitting density profiles and obtaining concentration measurements for typical realistic halos.  Large substructure, as displayed by the \za\ halo, can disrupt the radial symmetry of the halo and cause significant deviations in the density profile.  Centering can also be an issue in these cases.  Due to these complications, there are a number of approaches for finding halo concentrations \citep{2012MNRAS.423.3018P}, but for consistency, we use the values derived from \rockstar's fitting for our concentration measurements.

%~~~~~~~~~~~~~~~~~~~~~~~~~~~~~~~~~~~~~~~~~~~~~~~~~~~~~~~~~~~~~~~~~~~~~~~~~~~~~~~
% Histograms and Curve Fitting
%~~~~~~~~~~~~~~~~~~~~~~~~~~~~~~~~~~~~~~~~~~~~~~~~~~~~~~~~~~~~~~~~~~~~~~~~~~~~~~~

At each simulation snapshot, we measure and compare a number of parameters for halos in both \lpt\ and \za\ simulations.  For each quantity $q$, we create histograms of $\Delta q$, the normalized difference in $q$ between halos in the \lpt\ and \za\ simulations:
\begin{equation} \label{eq:delta_q}
	\Delta q = \frac{q_{\lpt} - q_{\za}}{q_{\mathrm{avg}}},
\end{equation}
where $q_{\mathrm{avg}} = \frac{1}{2} (q_{\lpt} + q_{\za})$.  The choice of $q_{\mathrm{avg}}$ for normalization allows us to be unbiased in our assumption of which halo better represents the truth, but can mask large differences between individual halos.  We fit each of these $\Delta q$ histograms with a generalized normal distribution \citep{doi:10.1080/02664760500079464} with the probability density function
\begin{equation} \label{eq:generalized_normal}
	f(x) = \frac{ \beta }{2 \alpha \Gamma(1 / \beta)} e^{\left( \left| x - \mu \right| / \alpha \right)^{\beta}},
\end{equation}
where $\mu$ is the mean, $\alpha$ is the scale parameter, $\beta$ is the shape parameter, and $\Gamma$ is the gamma function
\begin{equation} \label{eq:gamma_function}
	\Gamma(t) = \int_{0}^{\infty} x^{t-1} e^{-x} \dd x.
\end{equation}
The shape parameter $\beta$ is restricted to $\beta \geq 1$.  This allows the distribution to potentially vary from a Laplace distribution ($\beta = 1$) to a uniform distribution ($\beta = \infty$) and includes the normal distribution ($\beta = 2$).  The distribution has variance
\begin{equation} \label{eq:variance}
	\sigma^{2} = \frac{ \alpha^{2} \Gamma(3/\beta) }{ \Gamma(1/\beta) }
\end{equation}
and excess kurtosis
\begin{equation} \label{eq:kurtosis}
	\gamma_{2} = \frac{ \Gamma(5/\beta) \Gamma(1/\beta) }{ \Gamma(3/\beta)^{2} } - 3.
\end{equation}
The distribution is symmetric, and thus has no skewness by definition.  As such, the values for skew presented below are measured directly from the data.

As our fitting distributions are symmetrical, in order to derive uncertainties for skew, we measure the skew of the distributions for each of our three simulation boxes individually as well as for the single stacked data set.  Uncertainty in skew is then simply the standard deviation of the mean of the skew of the three individual boxes.

Determining the uncertainty in the kurtosis is slightly more involved, as kurtosis is determined by a transformation of the generalized normal distribution's shape parameter $\beta$ according to Equation~\ref{eq:kurtosis}.  Following the standard procedure for propagation of uncertainty, we calculate the standard deviation of the kurtosis:
\begin{align} \label{eq:kurt_err_partial}
    s_{\gamma_{2}} &= \sqrt{ \left( \frac{\diff \gamma_{2}}{\diff \beta} \right)^{2} s_{\beta}^{2} } \\
        &= s_{\beta} \frac{\diff}{\diff \beta} \left[ \frac{\Gamma(5/\beta) \Gamma(1/\beta)}{\Gamma(3/\beta)^{2}} - 3 \right].
\end{align}
The derivative of the gamma function is
\begin{equation} \label{eq:gamma_prime}
    \Gamma'(x) = \Gamma(x) \psi_{0}(x),
\end{equation}
where the digamma function $\digamma$ is the derivative of the logarithm of the gamma function and is given by
\begin{equation} \label{eq:digamma}
    \digamma(x) = \int_{0}^{\infty} \left( \frac{e^{-t}}{t} - \frac{e^{-xt}}{1 - e^{-t}} \right) \dd t
\end{equation}
if the real part of $x$ is positive.  Now, taking the derivative of $\gamma_{2}$ and doing a bit of algebra yields
\begin{equation} \label{eq:kurt_err}
    s_{\gamma_{2}} = s_{\beta} \frac{1}{\beta^{2}} \left( \gamma_{2} + 3 \right) \left[ 6 \digamma(3/\beta) - 5 \digamma(5/\beta) - \digamma(1/\beta) \right],
\end{equation}
with which we can find the uncertainty in the kurtosis given the value and uncertainty of the shape parameter $\beta$ estimated from the least squares fit routine.

In addition to distributions of $\Delta q$, we also consider distributions of
\begin{equation} \label{eq:delta_prime_q}
	\delta q = \frac{q_{\lpt} - q_{\za}}{q_{\za}}
\end{equation}
to better quantify the fraction of halos differing by a given amount between \lpt\ and \za\ simulations.  This is better suited to track the fractional differences between the halo populations and allows us to pose questions like:  how many \lpt\ halos are more massive than their \za\ counterparts by at least a given amount?  However, this function is inherently non-symmetrical, and is only defined for $\delta q \geq -1$ for positive quantities like mass and concentration.  Therefore, in order to count halo pairs that differ by a certain amount, regardless of whether $q$ is larger for the \lpt\ or \za\ halo, we define
\begin{equation} \label{eq:equivalent_q_prime}
	\delta q_{eq} = \frac{1}{\delta q + 1} - 1,
\end{equation}
the value for which a halo pair with a larger $q$ in \za\ would differ by the same factor as a halo pair with a larger $q$ in \lpt.

%%%%%%%%%%%%%%%%%%%%%%%%%%%%%%%%%%%%%%%%%%%%%%%%%%%%%%%%%%%%%%%%%%%%%%%%%%%%%%%%
%
% Results
%
%%%%%%%%%%%%%%%%%%%%%%%%%%%%%%%%%%%%%%%%%%%%%%%%%%%%%%%%%%%%%%%%%%%%%%%%%%%%%%%%

\section{Results}
\label{sec:results}

%%%%%%%%%%%%%%%%%%%%%%%%%%%%%%%%%%%%%%%%%%%%%%%%%%%%%%%%%%%%%%%%%%%%%%%%%%%%%%%%

With our catalog of matched dark matter halos, we directly compare differences in halo properties arising from initialization with \lpt\ vs \za.  We consider halos on a pair--by--pair basis as well as the entire sample as a whole.  Overall, we find \lpt\ halos have undergone more growth by a given redshift than their \za\ counterparts.

%~~~~~~~~~~~~~~~~~~~~~~~~~~~~~~~~~~~~~~~~~~~~~~~~~~~~~~~~~~~~~~~~~~~~~~~~~~~~~~~
\subsection{Individual halo pairs}
%~~~~~~~~~~~~~~~~~~~~~~~~~~~~~~~~~~~~~~~~~~~~~~~~~~~~~~~~~~~~~~~~~~~~~~~~~~~~~~~

We compare large scale morphologies, density profiles, and various other halo properties for halo pairs on an individual halo--by--halo basis for several of the most massive halos.  Morphologies appear similar for most halos, indicating good halo matches between simulations.  However, many pairs display differences in central morphology, such as the number and separation of central density peaks.  We interpret these cases to be examples of differences in merger epochs, in which case one halo may still be undergoing a major merger, while its companion is in a more relaxed post-merger state.  We give an example of one such pair at $z = 6$ in Figure~\ref{fig:halo-pair}.  The top two rows show density projections of the nuclear regions for a large \lpt\ and matching \za\ halo (first and second rows, respectively).  We find  the \za\ halo to contain two distinct density peaks with a separation of $\sim 10$ kpc, while the \lpt\ halo displays only a single core.  On the third and fourth rows, we plot the density profiles of the same two halos (\lpt\ and \za, respectively).  Here, with nearly identical virial radii, we can readily see that the \lpt\ halo is more concentrated than its \za\ counterpart.

%~~~~~~~~~~~~~~~~~~~~~~~~~~~~~~~~~~~~~~~~~~~~~~~~~~~~~~~~~~~~~~~~~~~~~~~~~~~~~~~
\subsection{Differences in ensemble halo properties}
%~~~~~~~~~~~~~~~~~~~~~~~~~~~~~~~~~~~~~~~~~~~~~~~~~~~~~~~~~~~~~~~~~~~~~~~~~~~~~~~

\begin{figure*}[t]
	\centering
	\begin{subfigure}{}
		\includegraphics[width=0.48\linewidth]{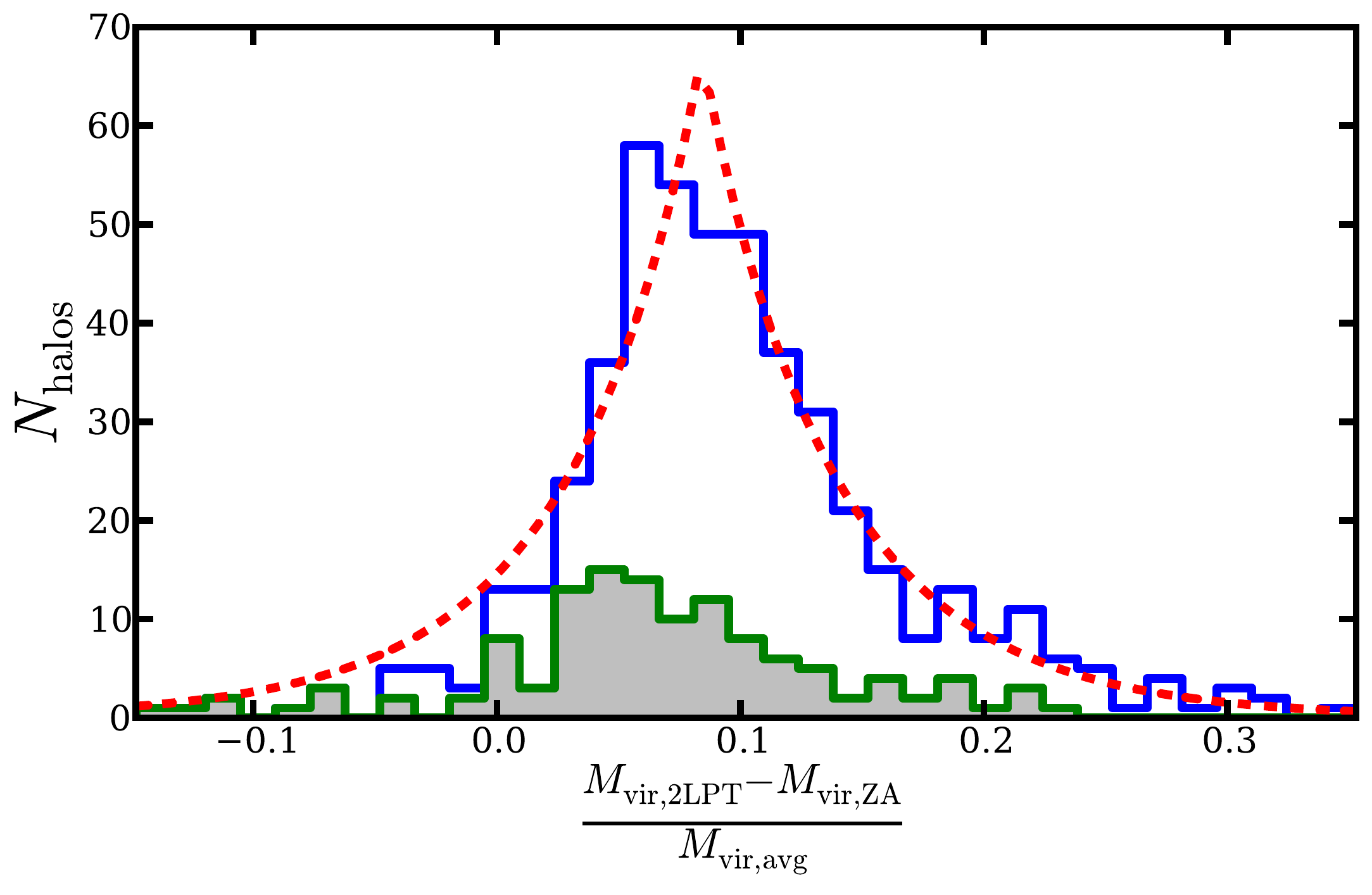}
	\end{subfigure}
	~
	\begin{subfigure}{}
		\includegraphics[width=0.48\linewidth]{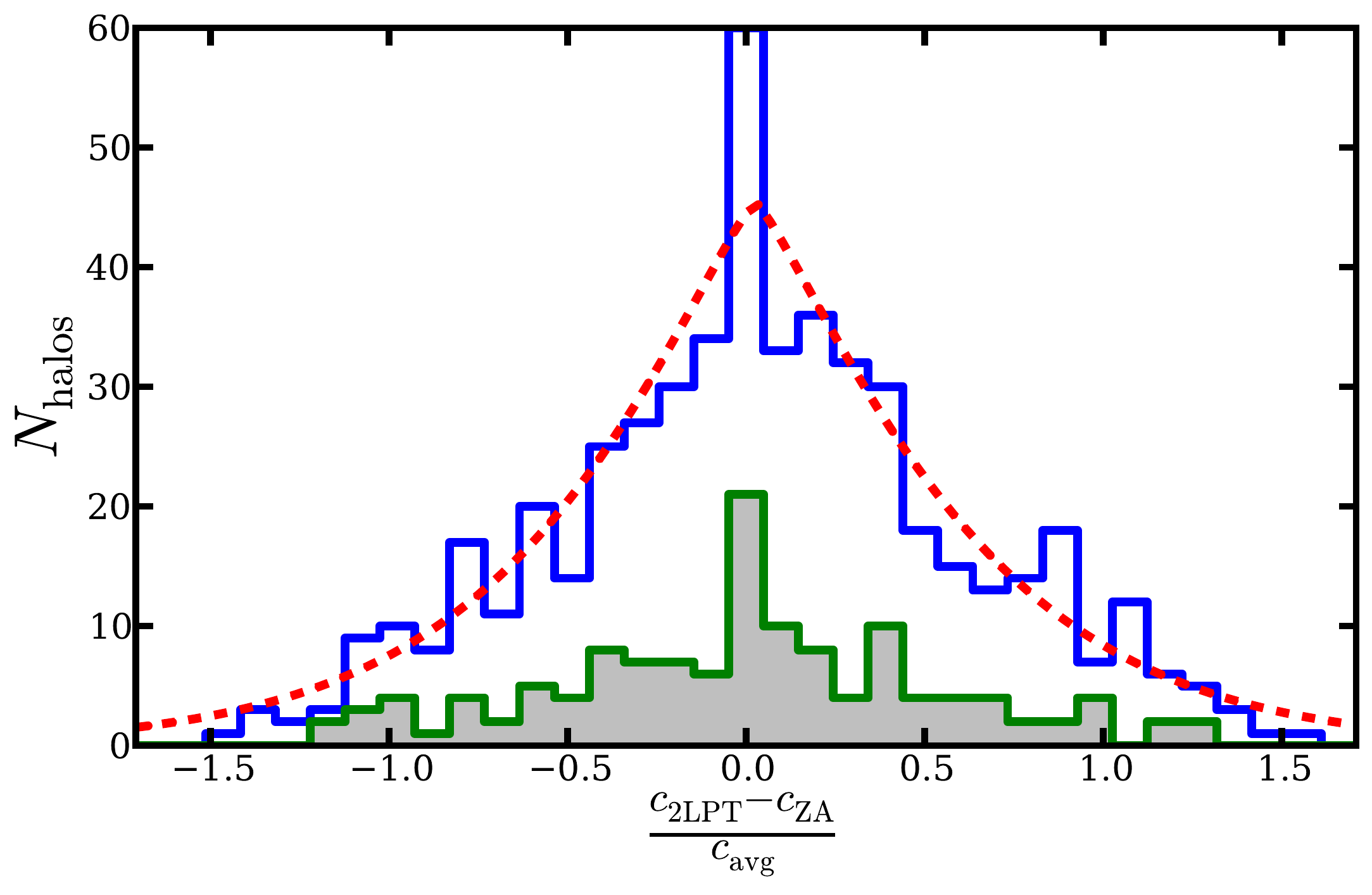}
	\end{subfigure}
	\\
	\begin{subfigure}{}
		\includegraphics[width=0.48\linewidth]{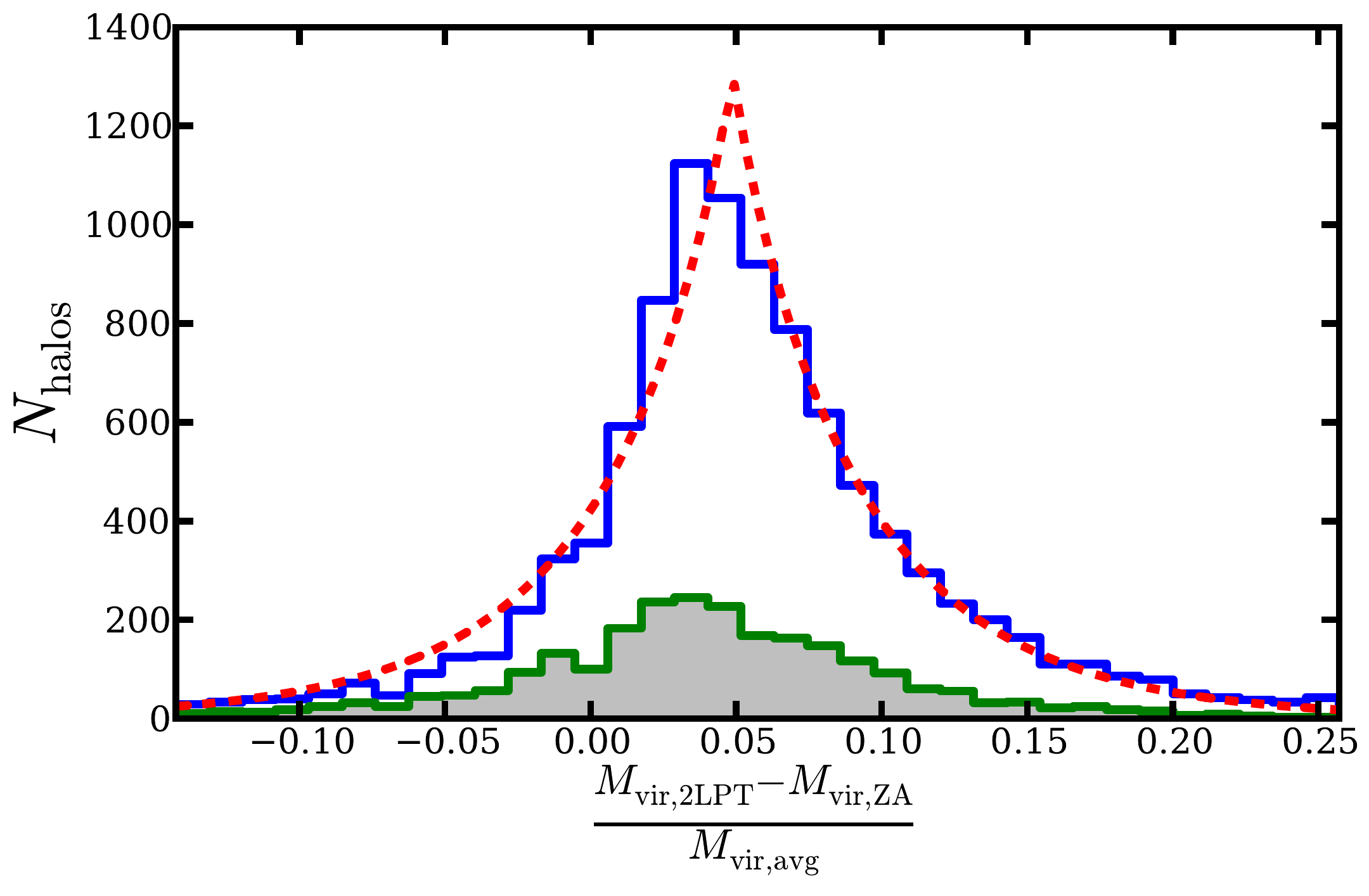}
	\end{subfigure}
	~
	\begin{subfigure}{}
		\includegraphics[width=0.48\linewidth]{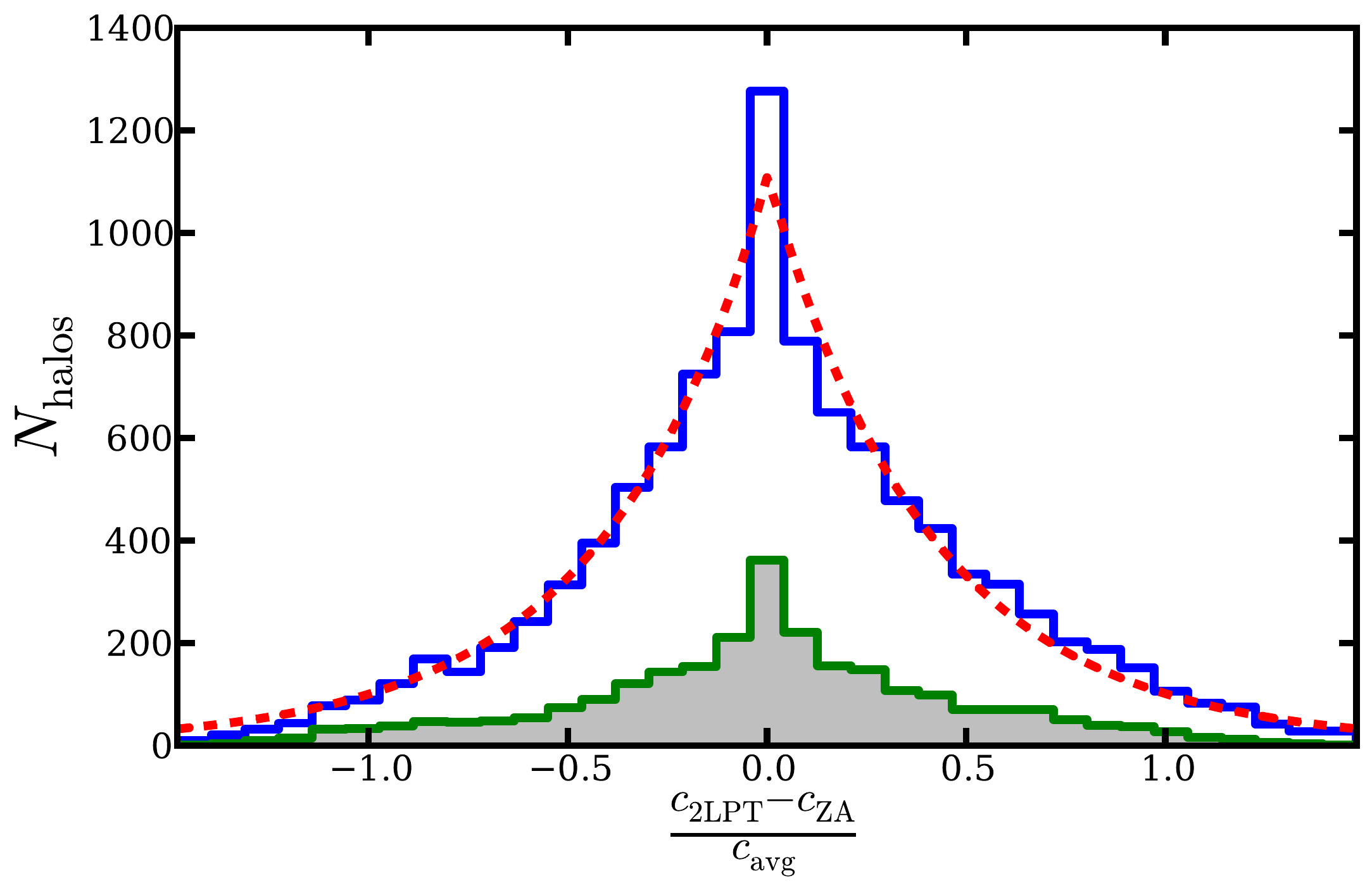}
	\end{subfigure}
	\\
	\begin{subfigure}{}
		\includegraphics[width=0.48\linewidth]{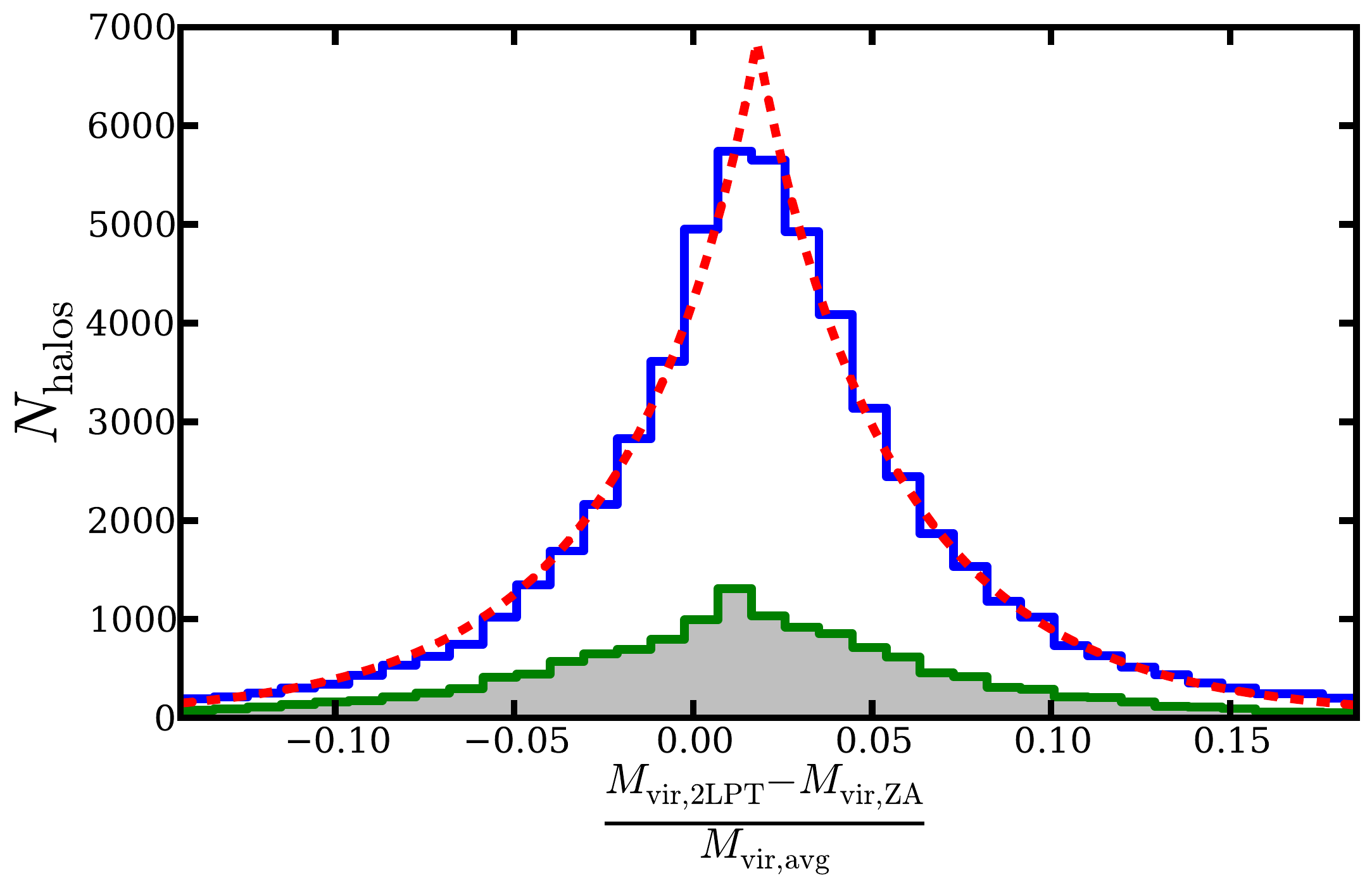}
	\end{subfigure}
	~
	\begin{subfigure}{}
		\includegraphics[width=0.48\linewidth]{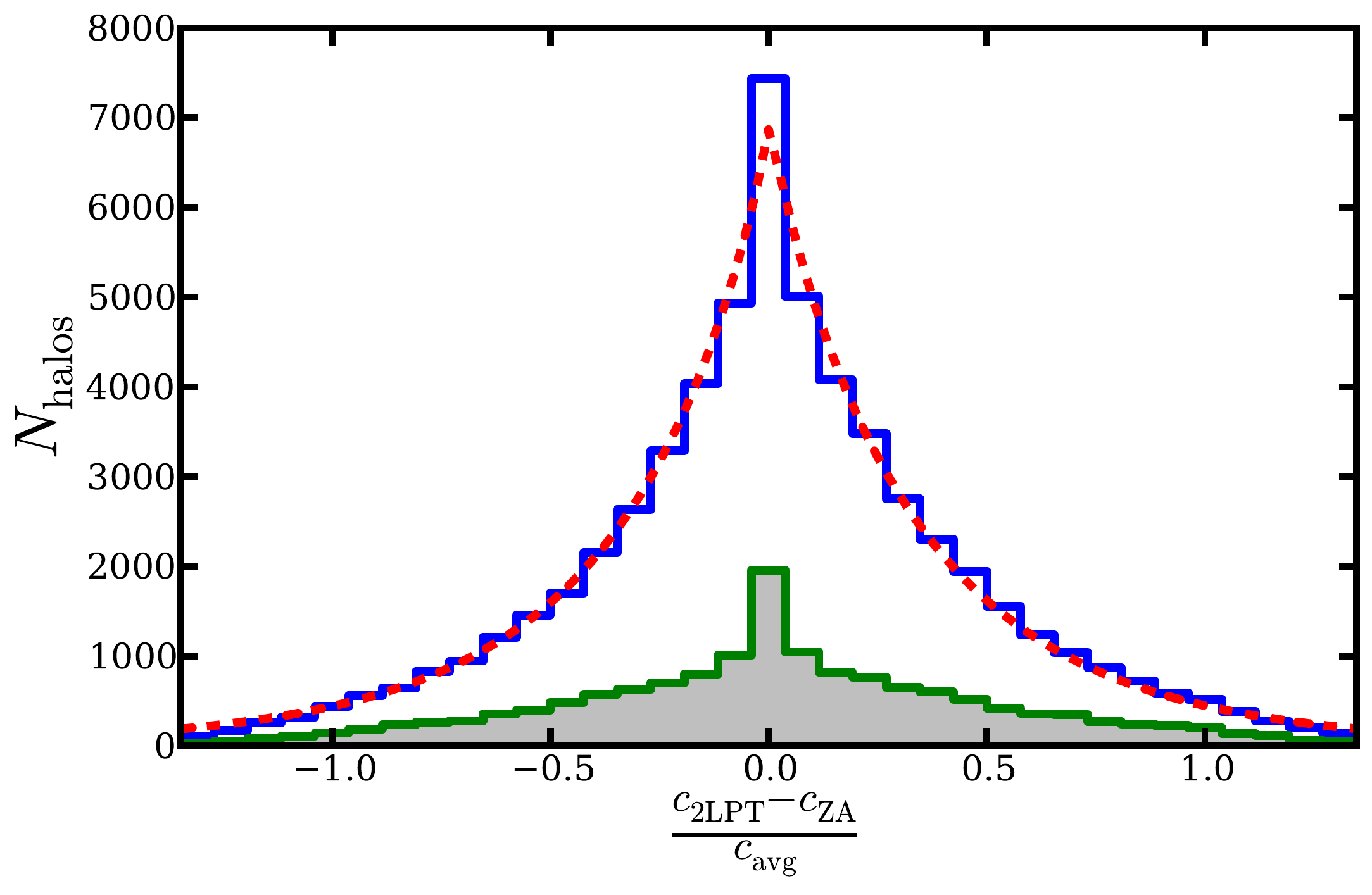}
	\end{subfigure}
	\caption[Histograms of $\Delta M_{\mathrm{vir}}$ and $\Delta c$]{\footnotesize Histograms of $\Delta M_{\mathrm{vir}}$ (\textit{left column}) and $\Delta c$ (\textit{right column}) for snapshots at $z = 14.7$, $z = 10.3$, and $z = 6.0$ (\textit{top, middle, and bottom panels, respectively}).  The small gray-filled histograms count only the top 25\% most massive halos.  The main histograms are fit with a generalized normal distribution, overplotted as red dashed curves, with parameters for mean, scale, and shape (see Equation~\ref{eq:generalized_normal}).  The distributions for $\Delta M_{\mathrm{vir}}$ have positive means and heavier \lpt\ halos, with the most pronounced difference at high redshift.  The distributions shown here have means of $(8.4 \pm 1.8) \times 10^{-2}$, $(4.87 \pm 0.87) \times 10^{-2}$, and $(1.79 \pm 0.31) \times 10^{-2}$, respectively.  The skew of the distribution is also the most positive at high redshift, and shifts toward symmetry by $z = 6$.  The $\Delta c$ distributions remain symmetric about zero and have negligible skew.  The means are consistent with zero, at $(2.6 \pm 2.7) \times 10^{-2}$, $(0.2 \pm 2.6) \times 10^{-2}$, and $(0.3 \pm 1.1) \times 10^{-2}$, respectively.  Both distributions have excess kurtosis consistently larger than that of a standard Gaussian distribution, with a sharp peak and heavy tails.}
	\label{fig:diff-hist}
\end{figure*}

For the halo population as a whole, we examine distributions of virial mass $M_{\mathrm{vir}}$ and concentration $c$.  We plot histograms of $\Delta M_{\mathrm{vir}}$ and $\Delta c$ in the left and right columns, respectively, of Figure~\ref{fig:diff-hist} for redshifts 14.7, 10.3, and 6.0.  For each panel, the blue histogram features the entire halo sample, and the smaller gray-filled green histogram displays only the top 25\% most massive halos, ordered by \lpt\ mass.  Fits to the primary histograms are overplotted as red dashed curves.

Throughout the simulation, we find a tendency for \lpt\ halos to be more massive.  At $z = 15$, the mean of the $\Delta M_{\mathrm{vir}}$ distribution is $(9.3 \pm 1.2) \times 10^{-2}$.  The mean is consistently positive (heavier \lpt\ halos) and is most displaced from zero at high redshift.  The peak of the distribution gradually moves closer to zero as we progress in redshift.  We find the least difference between paired halos for the final snapshot at $z = 6$, with $\mu_{\Delta M_{\mathrm{vir}}} = (1.79 \pm 0.31) \times 10^{-2}$.

The higher-order moments of the $\Delta M_{\mathrm{vir}}$ distribution are of interest as well, as we find significant deviation from a Gaussian distribution.  One may expect this from the non-linear nature of gravitational collapse; the most massive outliers collapse earlier in \lpt, and this head start compounds subsequent evolution.  As we use a symmetrical generalized normal distribution to fit the data, the skew cannot be recovered from the fit itself; we therefore measure deviation from symmetry directly from the data.  By $z = 6$, we observe a rather symmetrical distribution, with both sides of the histogram equally well described by our fit.  However, at higher redshift, we note a marked increase in skewness and deviation from this symmetry.  As redshift increases, we observe an increasing difference between the fit curve and the bins to the left of the histogram peak.

We find the distributions to be much closer to a Laplace distribution than a Gaussian, with shape parameter consistently sitting at or very close to $\beta = 1$.  Compared to a Gaussian distribution, the larger excess kurtosis implies a narrower central peak and heavier outlying tails.  Our fit is constrained such that $\beta \geq 1$, so the kurtosis of the data itself could potentially be higher than the fit implies.

We find no overall preference for more concentrated \lpt\ or \za\ halos.  In contrast to the $\Delta M_{\mathrm{vir}}$ histograms, $\Delta c$ shows very little deviation from symmetry about zero.  Throughout the simulation, we find the distributions to have a mean close to zero and negligible skew.  The widths of the distributions are much larger than those for $\Delta M_{\mathrm{vir}}$, with the standard deviation of the $\Delta c$ distributions consistently about an order of magnitude higher than for $\Delta M_{\mathrm{vir}}$.  As with mass, concentration histograms are sharply peaked with heavy tails, implying a tendency for halo pairs to move towards the extremes of either very similar or very discrepant concentrations.

%:::::::::::::::::::::::::::::::::::::::::::::::::::::::::::::::::::::::::::::::
\subsubsection{Time evolution of mass and concentration differences}
%:::::::::::::::::::::::::::::::::::::::::::::::::::::::::::::::::::::::::::::::

\begin{figure*}[t]
	\centering
	\begin{subfigure}{}
		\includegraphics[width=0.48\linewidth]{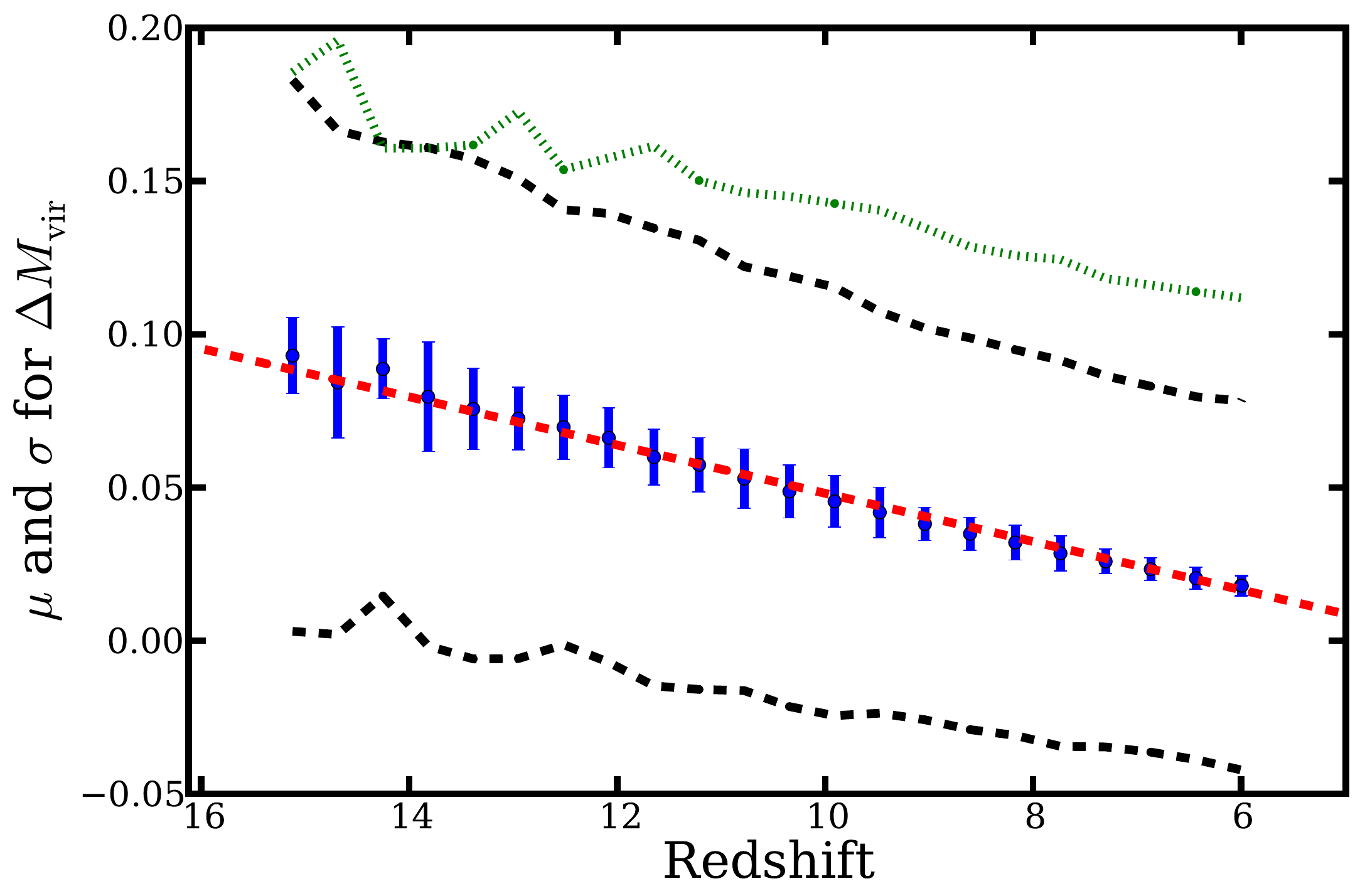}
	\end{subfigure}
	~
	\begin{subfigure}{}
		\includegraphics[width=0.48\linewidth]{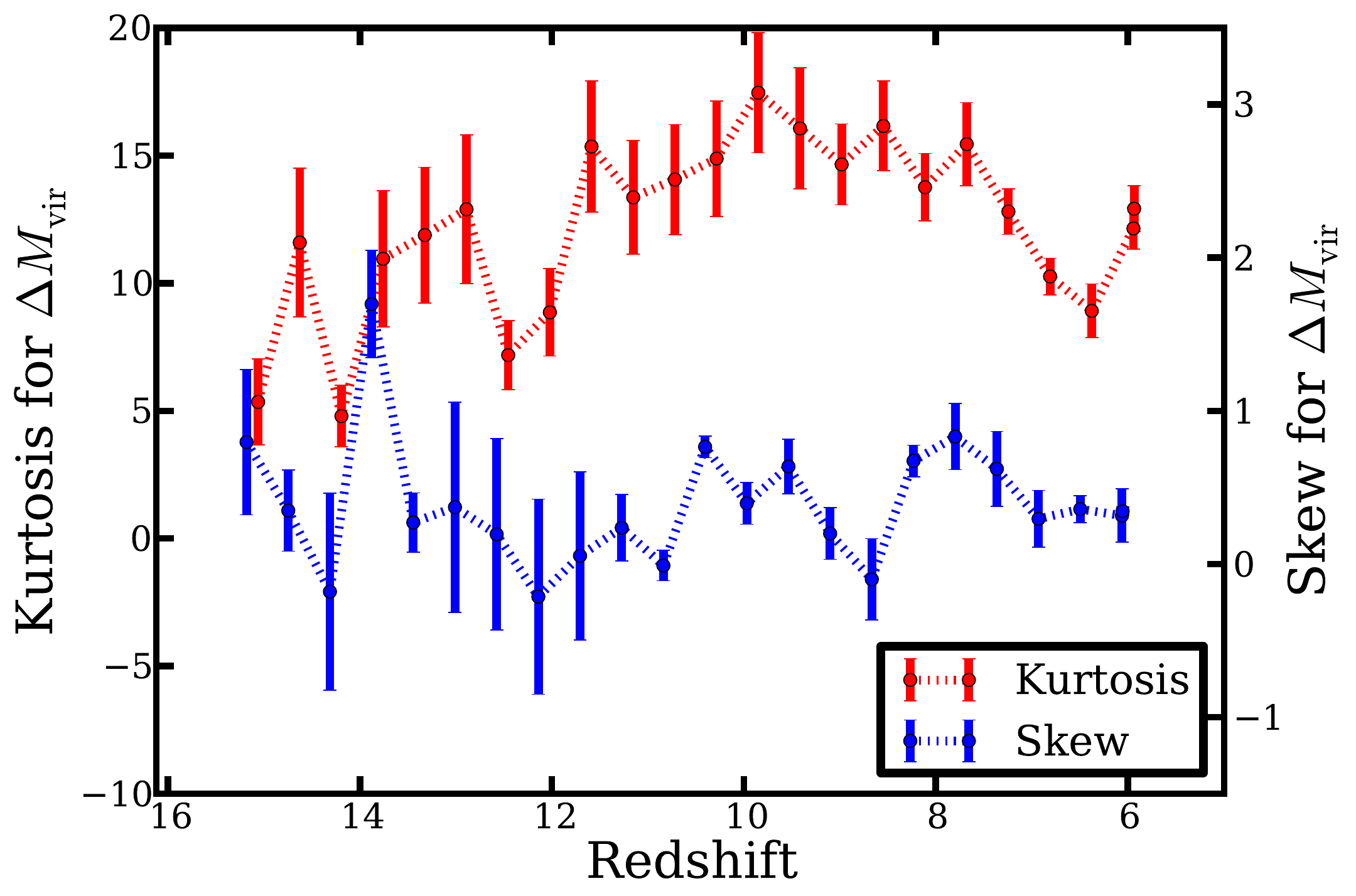}
	\end{subfigure}
	\\
	\begin{subfigure}{}
		\includegraphics[width=0.48\linewidth]{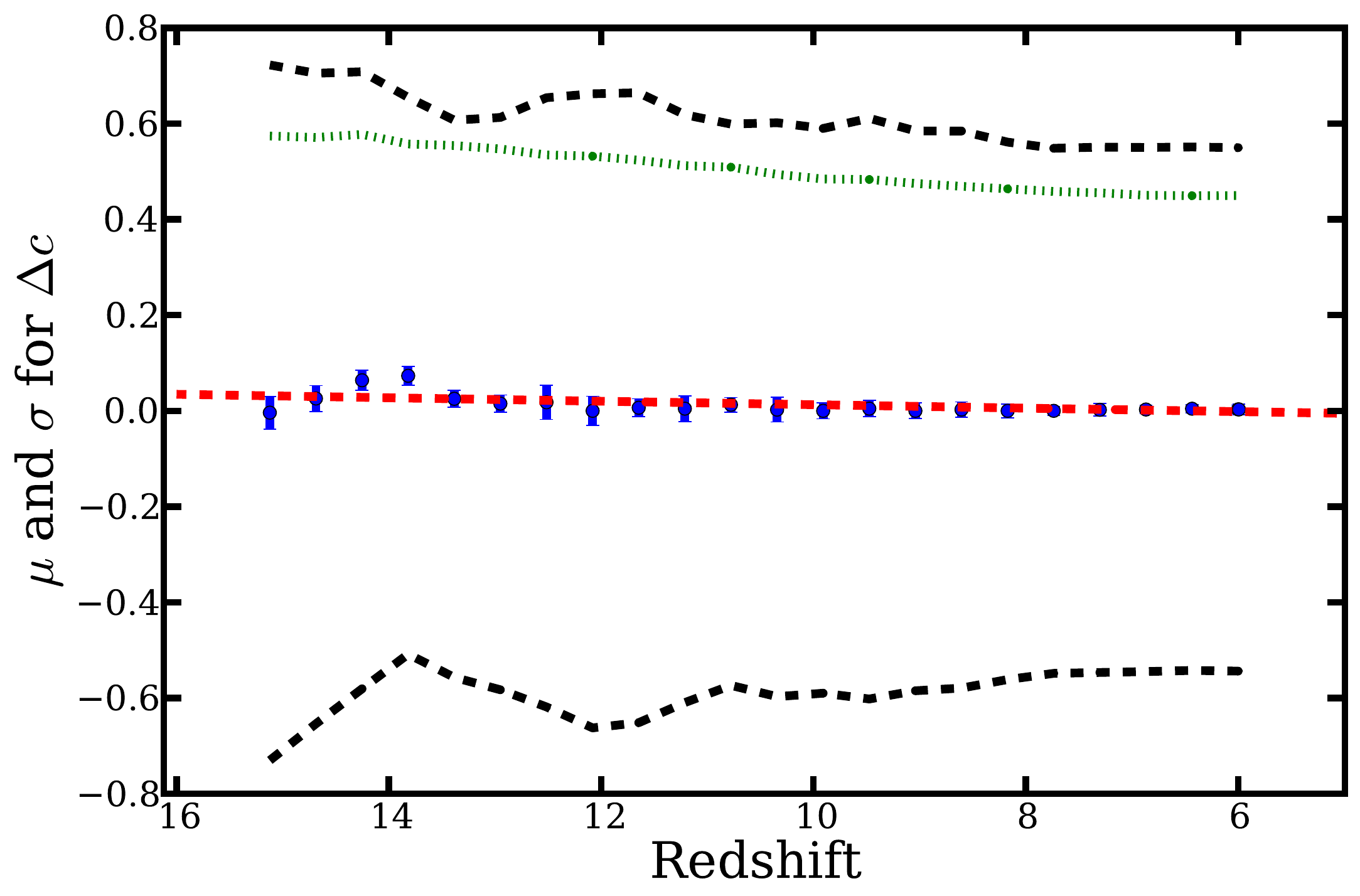}
	\end{subfigure}
	~
	\begin{subfigure}{}
		\includegraphics[width=0.48\linewidth]{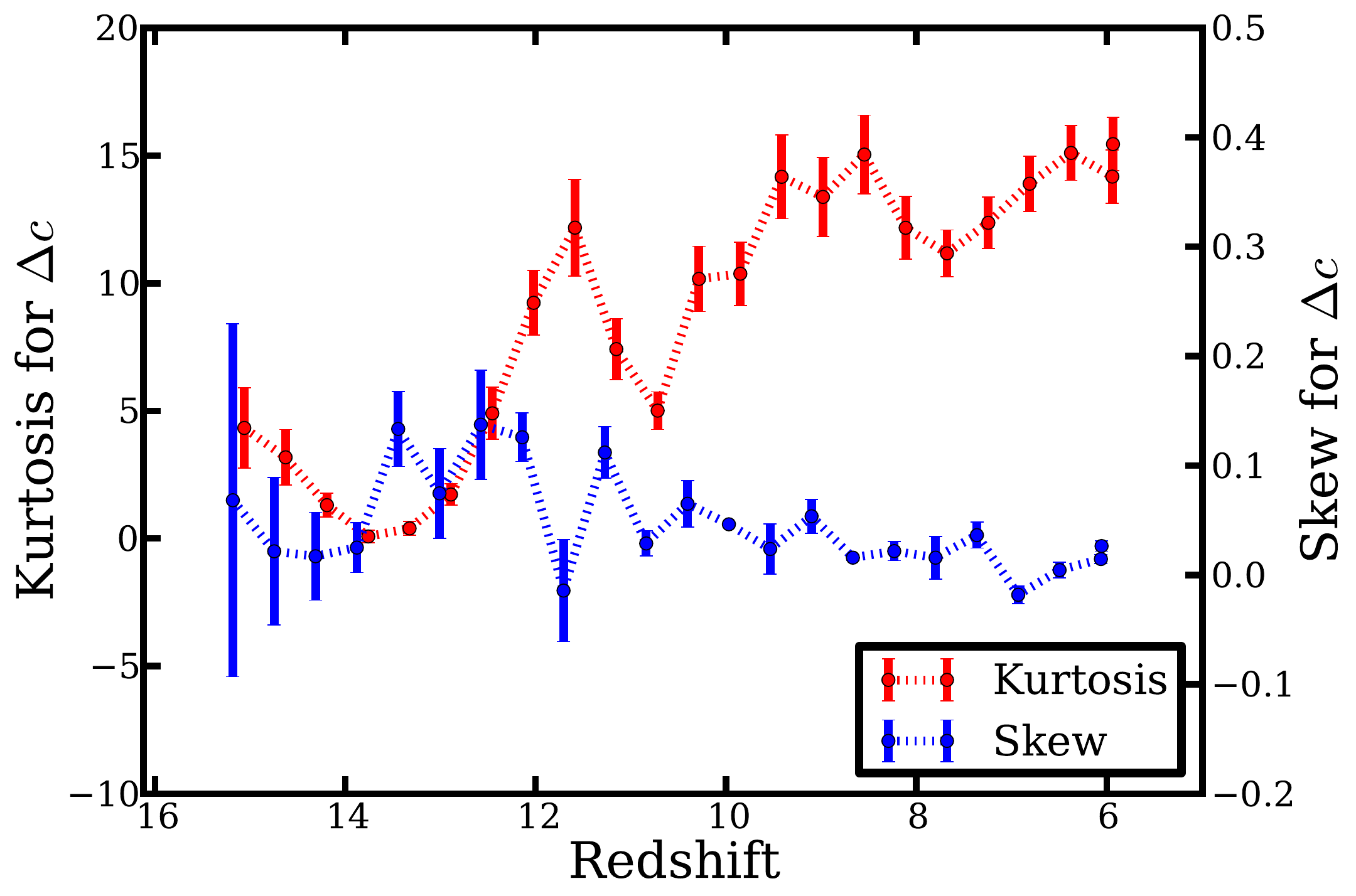}
	\end{subfigure}
	\caption[Statistics as functions of redshift for generalized normal fits]{\footnotesize Mean, standard deviation, and rms (\textit{left column}) and skew and excess kurtosis (\textit{right column}) as functions of redshift for $\Delta M_{\mathrm{vir}}$ (\textit{top row}) and $\Delta c$ (\textit{bottom row}).  In the left column, $\mu$ is plotted as blue points, $\mu \pm \sigma$ is plotted as the black dashed curves, and rms values are plotted as a green dotted curve.  The red dashed line is a linear fit to the mean.  We find a significant trend for $\mu$ for $\Delta M_{\mathrm{vir}}$ to be more positive at higher redshift and gradually shift toward zero as the simulation progresses, with a fit function of $\mu_{\Delta M_{\mathrm{vir}}} = (7.88 \pm 0.17) \times 10^{-3} z - (3.07 \pm 0.14) \times 10^{-2}$.  The mean for $\Delta c$, however, remains at or very near zero for most of the simulation and is fit by $\mu_{\Delta c} = (3.62 \pm 0.95) \times 10^{-3} z - (2.34 \pm 0.84) \times 10^{-2}$.  The $\Delta M_{\mathrm{vir}}$ and $\Delta c$ distributions narrow over time, with a slight decrease in $\sigma$.  In the right column, we plot skew (blue curve) and excess kurtosis (red curve).  Skew is positive for much of the simulation for $\Delta M_{\mathrm{vir}}$, but is much smaller for $\Delta c$.  Kurtosis is large (much more peaked than Gaussian) for both $\Delta M_{\mathrm{vir}}$ and $\Delta c$ throughout much of the simulation, and especially at later redshift.}
	\label{fig:fit_trends}
\end{figure*}

In Figure~\ref{fig:fit_trends}, we more quantitatively assess the evolution of our various trends hinted at in Figure~\ref{fig:diff-hist}.  Here, we plot the mean, root mean square (rms), standard deviation, skew, and kurtosis for $\Delta M_{\mathrm{vir}}$ and $\Delta c$ as functions of redshift.  Uncertainty in the mean is estimated directly from least squares theory.

The mean for $\Delta M_{\mathrm{vir}}$ is positive and highest at high redshift, trending toward zero by the end of the simulation.  Distributions for $\Delta c$ retain means close to and consistent with zero.  Standard deviation decreases slightly for both $\Delta M_{\mathrm{vir}}$ and $\Delta c$.  From $z = 15$ to $z = 6$, standard deviation falls from $(9.0 \pm 1.5) \times 10^{-2}$ to $(6.08 \pm 0.31) \times 10^{-2}$ for $\Delta M_{\mathrm{vir}}$ and from $0.73 \pm 0.11$ to $0.551 \pm 0.026$ for $\Delta c$.

\begin{table}[t]
	\centering
	\caption{Coefficients for linear least squares fits from Figure~\ref{fig:fit_trends}.}
	\begin{tabular}{ c  r  r }
		\toprule
		                           &  \multicolumn{1}{c}{$A$}             &  \multicolumn{1}{c}{$B$} \\
		\cmidrule(l){2-3}
		$\Delta M_{\mathrm{vir}}$  &  $(7.88 \pm 0.17) \times 10^{-3}$  &  $(-3.07 \pm 0.14) \times 10^{-2}$ \\
		$\Delta c$                 &  $(3.62 \pm 0.95) \times 10^{-3}$  &  $(-2.34 \pm 0.84) \times 10^{-2}$ \\
		\bottomrule
	\end{tabular}
	\label{tab:coeffs}
\end{table}

We find least square linear fits for both mean $\Delta M_{\mathrm{vir}}$ vs $z$ and mean $\Delta c$ vs $z$.  Coefficients for slope $A$ and y-intercept $B$ for the fit equation $\mu = A z + B$ are given in Table~\ref{tab:coeffs} for both cases.  We find a significant trend for $\Delta M_{\mathrm{vir}}$, with a slope $\sim 46 \sigma$ from zero.  Conversely, the slope for $\Delta c$ is much smaller and, considering the larger spread of the underlying distributions, can be considered negligible.  For $\Delta M_{\mathrm{vir}}$, the y-intercept coefficient $B$ likely has little meaning in terms of the actual behavior at $z = 0$, as we expect the trend to level out at later redshift.

We do note, however, that the mean can be deceiving as an indicator of total difference between halo populations, especially when it is close to zero as with concentration.  It should be noted that while the mean can indicate a lack of average difference between the whole sample of \lpt\ and \za\ halos, there can still be very large discrepancies between many individually paired halos.  We visualize this by plotting the rms of $\Delta M_{\mathrm{vir}}$ and $\Delta c$, which is plotted as a green dotted curve.  Unlike the mean, standard deviation, and kurtosis, which are measured from fits to the histograms, rms is measured directly from the data and is not dependent on fitting.  The large rms values are indicative of how much overall difference can arise between \lpt\ and \za\ halos, even though the differences may average to zero when considering the entire population.  The rms for both $\Delta M_{\mathrm{vir}}$ and $\Delta c$ starts highest at high redshift---$0.19$ for $\Delta M_{\mathrm{vir}}$ and $0.57$ for $\Delta c$ at $z = 15$---and steadily decreases throughout the simulation, reaching minimums of $0.11$ for $\Delta M_{\mathrm{vir}}$ and $0.45$ for $\Delta c$ by $z = 6$.

Additionally, it is of interest to consider the percentage of halo pairs that are ``wrong'' at some given time, regardless of whether the quantity is higher in \lpt\ or \za.  For example, if we count halos outside a slit of $\epsilon = 10\%$ around $\Delta q = 0$, we find that by $z = 6$, 14.6\% of halo pairs still have substantially mismatched masses, and 74.3\% have mismatched concentrations.  It is evident that a substantial percentage of halo pairs can have markedly different growth histories, even when there is little or no offset in the ensemble halo population average.

Kurtosis is consistently large for both mass and concentration, with a slight increasing trend throughout the simulation for concentration.  It reaches maximum values of $17.5 \pm 2.4$ at redshift 10 for $\Delta M_{\mathrm{vir}}$ and $15.4 \pm 1.0$ at the end of the simulation at redshift 6 for $\Delta c$.  Skew is positive for much of the simulation for mass, but is much smaller for concentration.  We find average skews of $0.39 \pm 0.29$ for $\Delta M_{\mathrm{vir}}$ and $0.045 \pm 0.028$ for $\Delta c$.  These higher moment deviations from Gaussianity again hint at the non-linear dynamics at play in halo formation.

The narrow peak and heavy tails of the distribution may indicate a fair amount of sensitivity to initial differences in halo properties, in that halo pairs that start out within a certain range of the mean are more likely to move closer to the mean, while pairs that are initially discrepant will diverge even further in their characteristics.  This is indicative of the non-linear gravitational influence present during halo evolution, and is further supported by a kurtosis that increases with time.

The skew at high redshift for $\Delta M_{\mathrm{vir}}$ may give another hint at the non-linear halo formation process.  Runaway halo growth causes more massive halos to favor even faster mass accretion and growth.  The positively skewed distributions show a picture of \lpt\ halo growth in which initial differences in mass are amplified most readily in the earliest forming and most massive halos, again indicating the extra kick-start to halo growth provided by \lpt\ initialization.  While the slight decrease in skew with redshift may be counter-intuitive to this notion, it is likely that the large number of newly formed halos begin to mask the signal from the smaller number of large halos displaying this effect.

%:::::::::::::::::::::::::::::::::::::::::::::::::::::::::::::::::::::::::::::::
\subsubsection{Global halo population differences as a function of  halo mass}
%:::::::::::::::::::::::::::::::::::::::::::::::::::::::::::::::::::::::::::::::

\begin{figure*}[t]
	\centering
	\begin{subfigure}{}
		\includegraphics[width=0.48\linewidth]{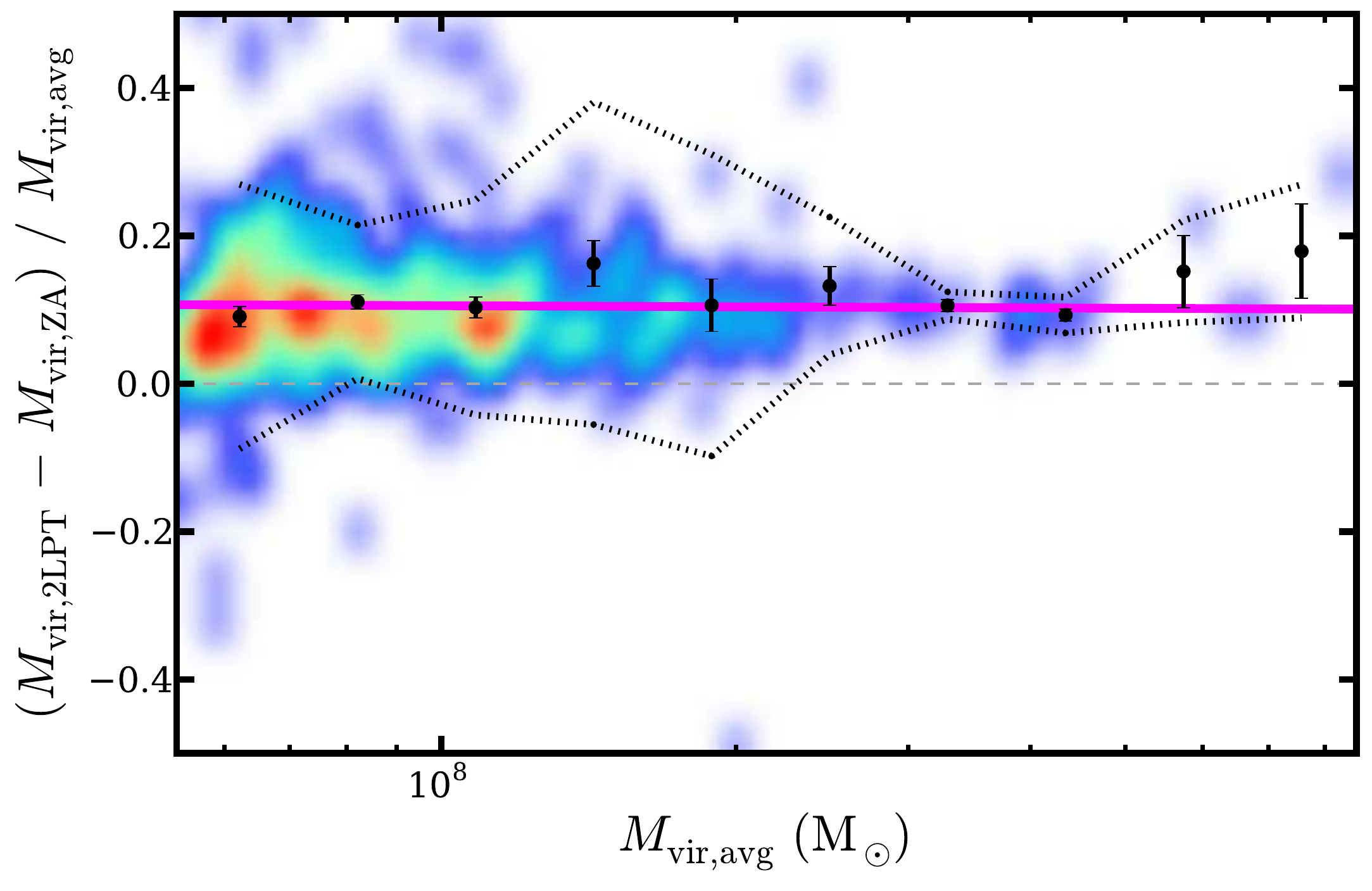}
	\end{subfigure}
	~
	\begin{subfigure}{}
		\includegraphics[width=0.48\linewidth]{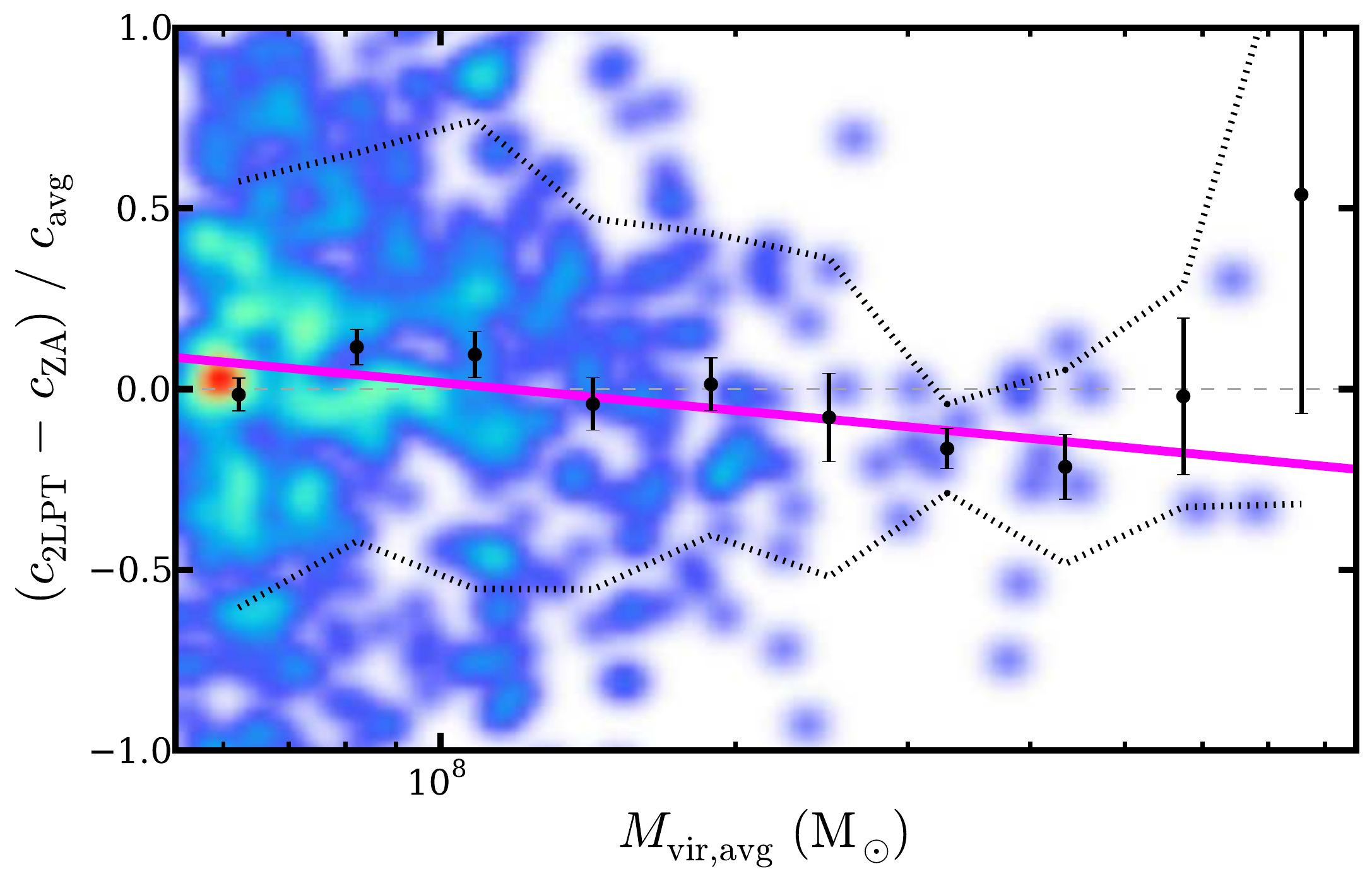}
	\end{subfigure}
	\\
	\begin{subfigure}{}
		\includegraphics[width=0.48\linewidth]{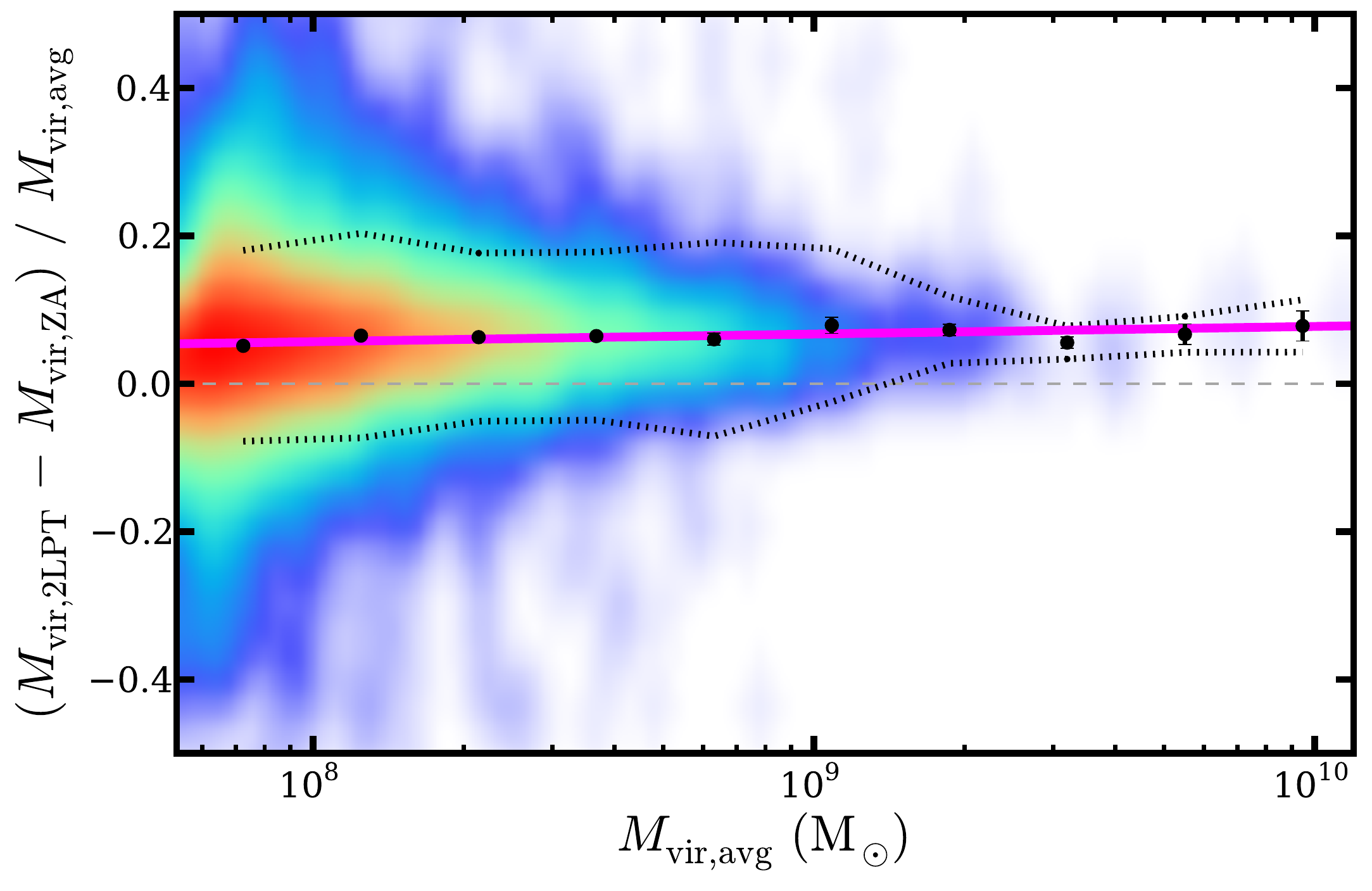}
	\end{subfigure}
	~
	\begin{subfigure}{}
		\includegraphics[width=0.48\linewidth]{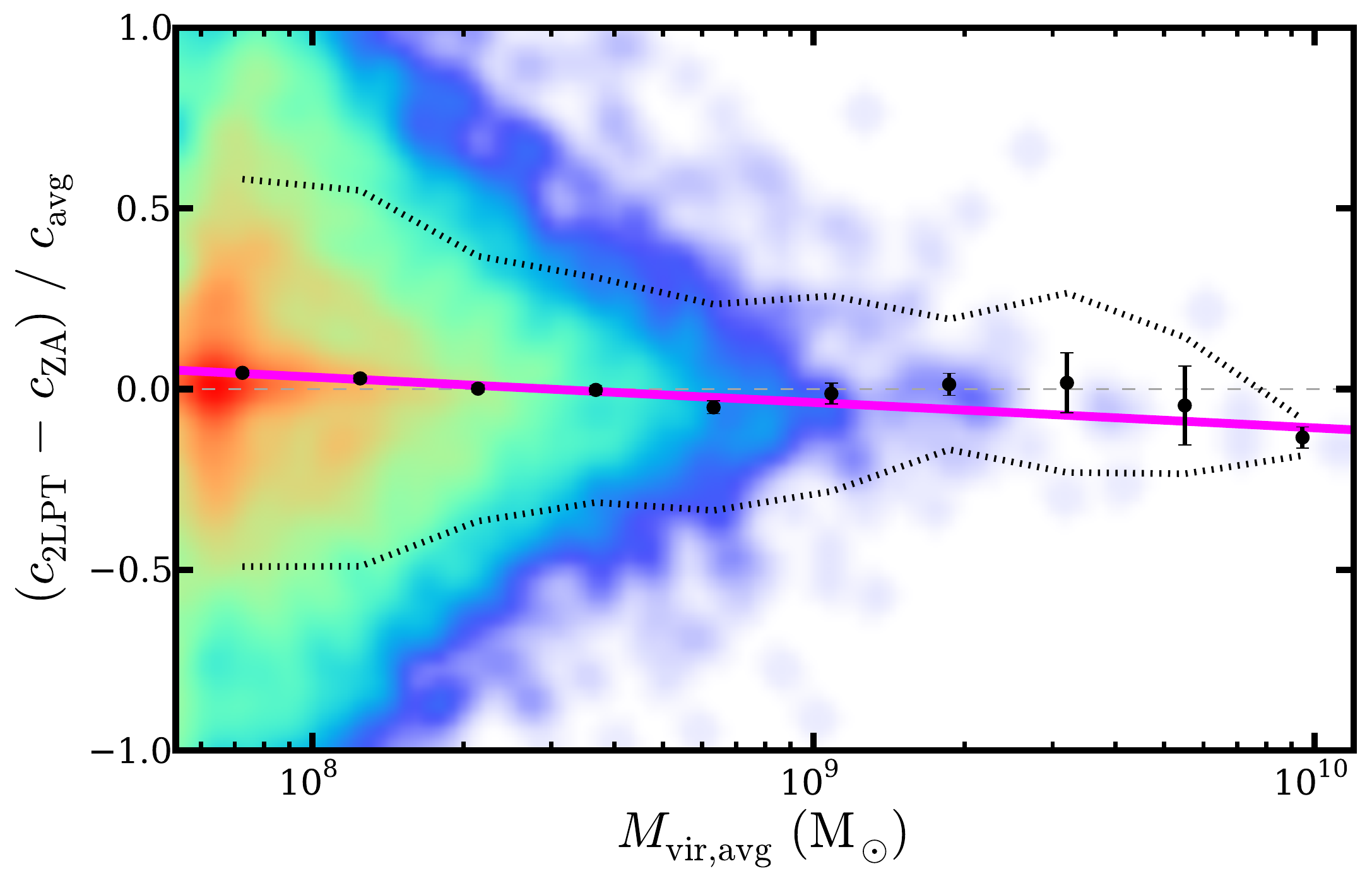}
	\end{subfigure}
	\\
	\begin{subfigure}{}
		\includegraphics[width=0.48\linewidth]{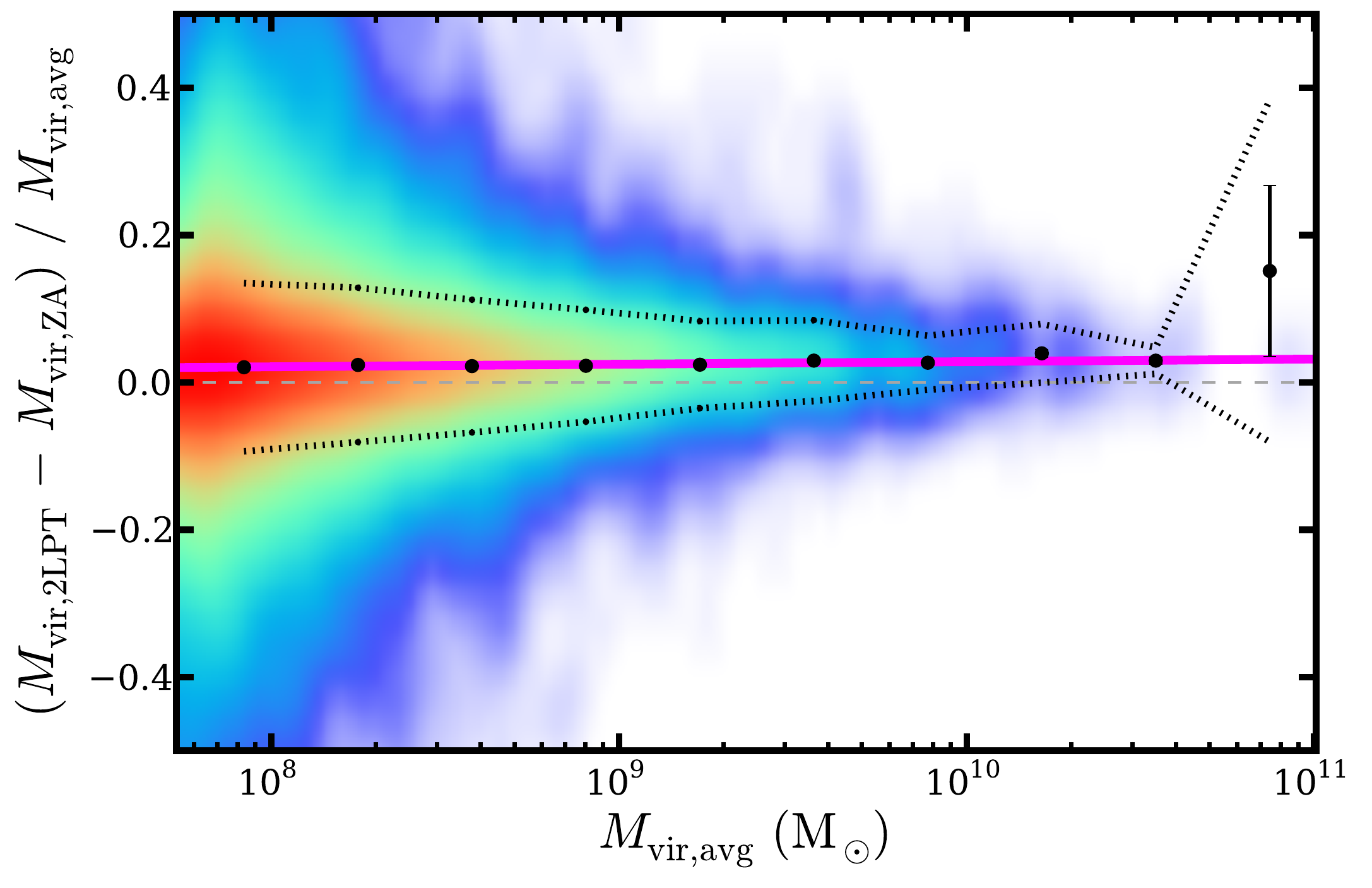}
	\end{subfigure}
	~
	\begin{subfigure}{}
		\includegraphics[width=0.48\linewidth]{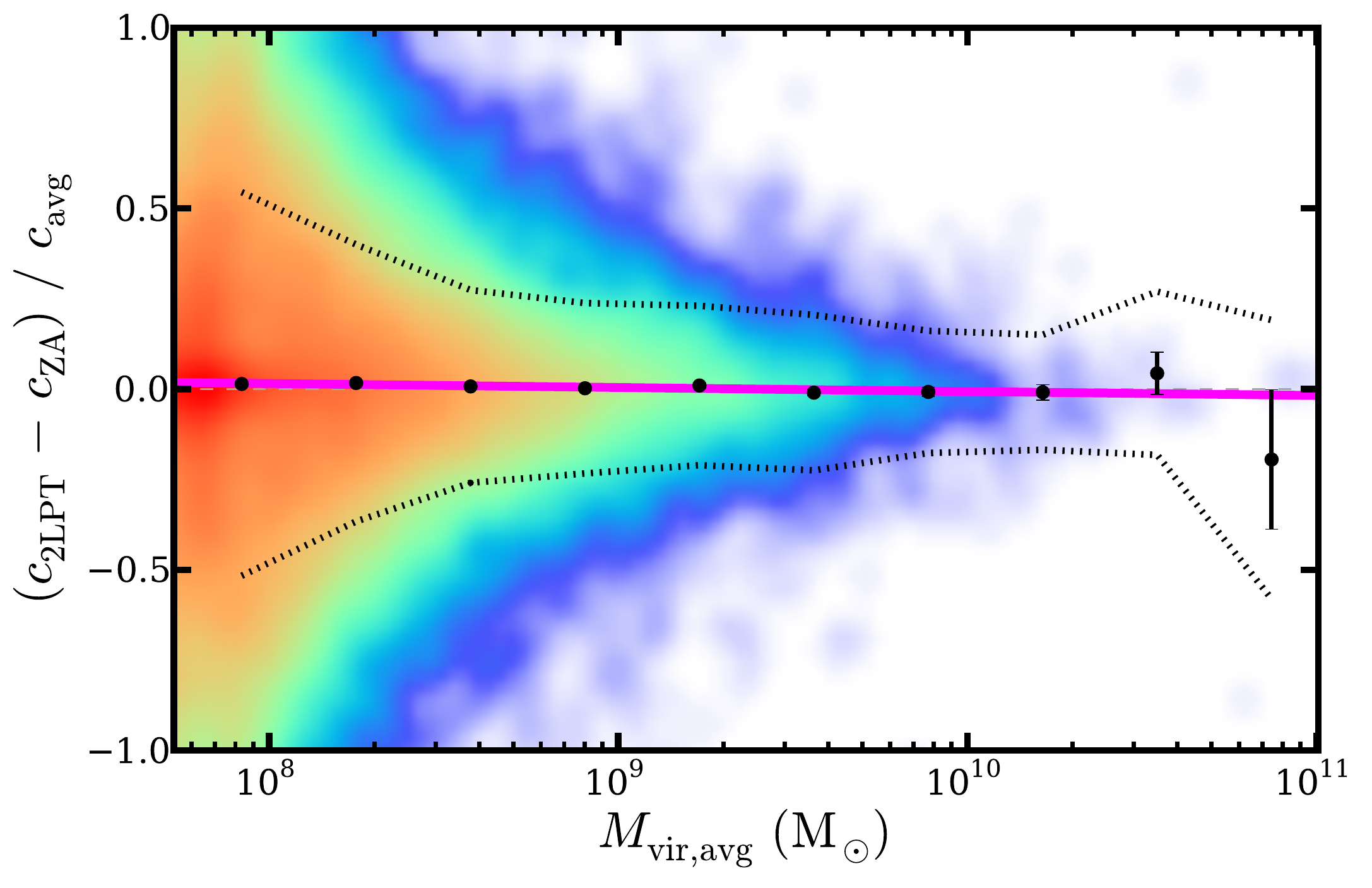}
	\end{subfigure}
	\caption[$\Delta M_{\mathrm{vir}}$ and $\Delta c$ as a function of $M_{\mathrm{vir,avg}}$]{\footnotesize $\Delta M_{\mathrm{vir}}$ (left column) and $\Delta c$ (right column) as functions of $M_{\mathrm{vir,avg}}$.  For the 2-D color histogram, halos are counted in rectangular bins and smoothed with a Gaussian kernel with a logarithmic color scale.  The halos are also divided into logarithmically-spaced bins in average virial mass, and the mean for each bin is plotted as a black point.  The black dotted curves are the standard deviation around the mean.  The magenta line is the linear least-squares best fit to the bin means.  The light grey dashed line at $\Delta q = 0$ is provided to guide the eye.  The three rows again correspond to snapshots at $z = 14.7$, $z = 10.3$, and $z = 6.0$.  We again see the overall offset for positive $\Delta M_{\mathrm{vir}}$ as before, and additionally find a small tendency for more massive halo pairs to be more likely to have even larger $\Delta M_{\mathrm{vir}}$.  Fit equations for the left column panels are $\Delta M_{\mathrm{vir}} = -(0.5 \pm 1.5) \times 10^{-2} \log(M_{\mathrm{vir,avg}}) + (0.15 \pm 0.12)$, $\Delta M_{\mathrm{vir}} = (1.03 \pm 0.46) \times 10^{-2} \log(M_{\mathrm{vir,avg}}) - (2.6 \pm 3.8) \times 10^{-2}$, and $\Delta M_{\mathrm{vir}} = (3.49 \pm 0.99) \times 10^{-3} \log(M_{\mathrm{vir,avg}}) - (6.8 \pm 8.3) \times 10^{-3}$, respectively.  Concentration shows an opposite trend where more massive halos are less concentrated in \lpt\ than in \za.  The right column panels have fit equations $\Delta c = -(0.256 \pm 0.093) \log(M_{\mathrm{vir,avg}}) + (2.07 \pm 0.76)$, $\Delta c = -(7.0 \pm 1.2) \times 10^{-2} \log(M_{\mathrm{vir,avg}}) + (0.595 \pm 0.099)$, and $\Delta c = -(1.10 \pm 0.31) \times 10^{-2} \log(M_{\mathrm{vir,avg}}) + (0.103 \pm 0.026)$, respectively.}
	\label{fig:delta-v-Mavg}
\end{figure*}

\begin{figure*}[t]
	\centering
	\begin{subfigure}{}
		\includegraphics[width=0.48\linewidth]{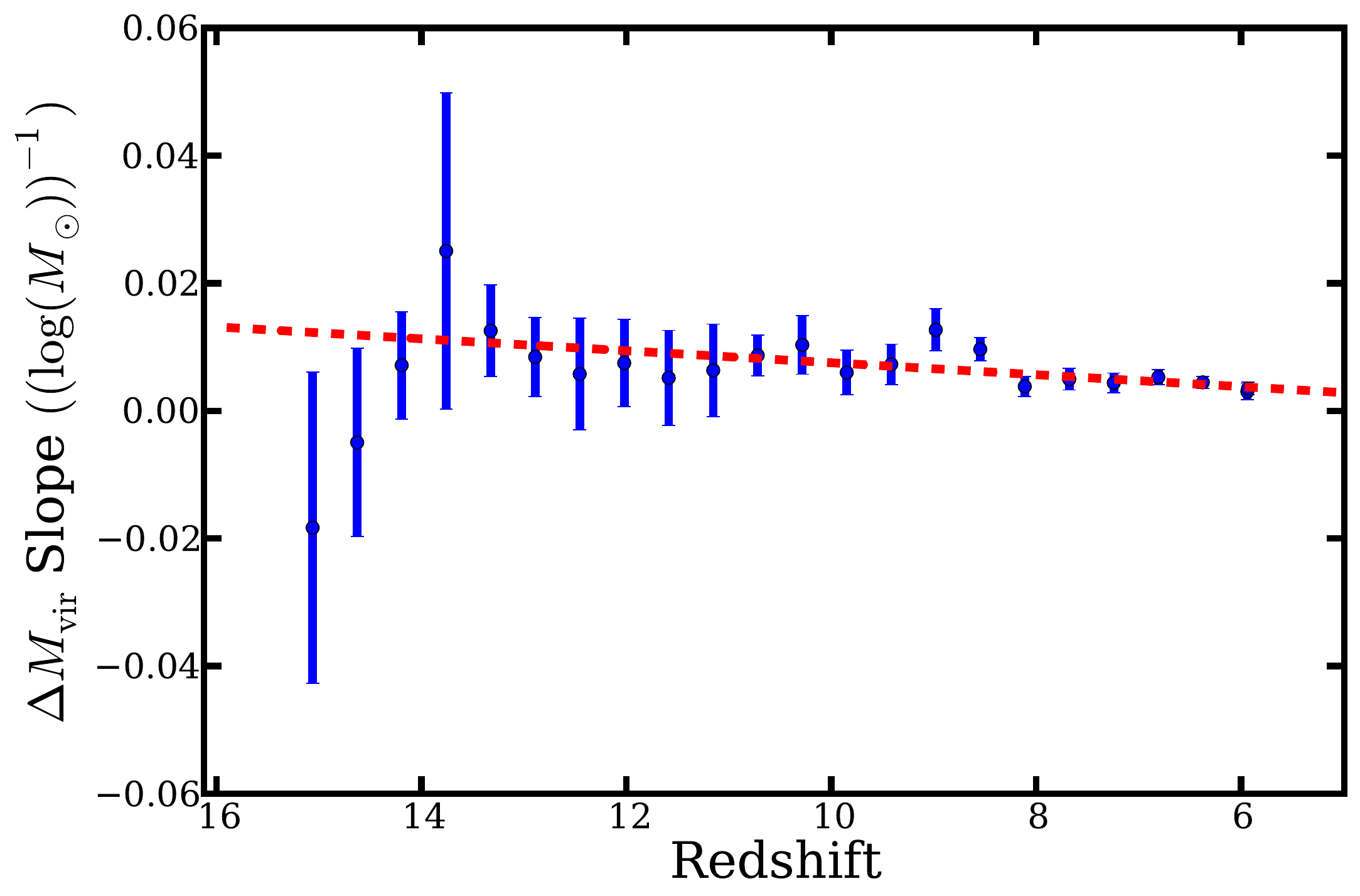}
	\end{subfigure}
	~
	\begin{subfigure}{}
		\includegraphics[width=0.48\linewidth]{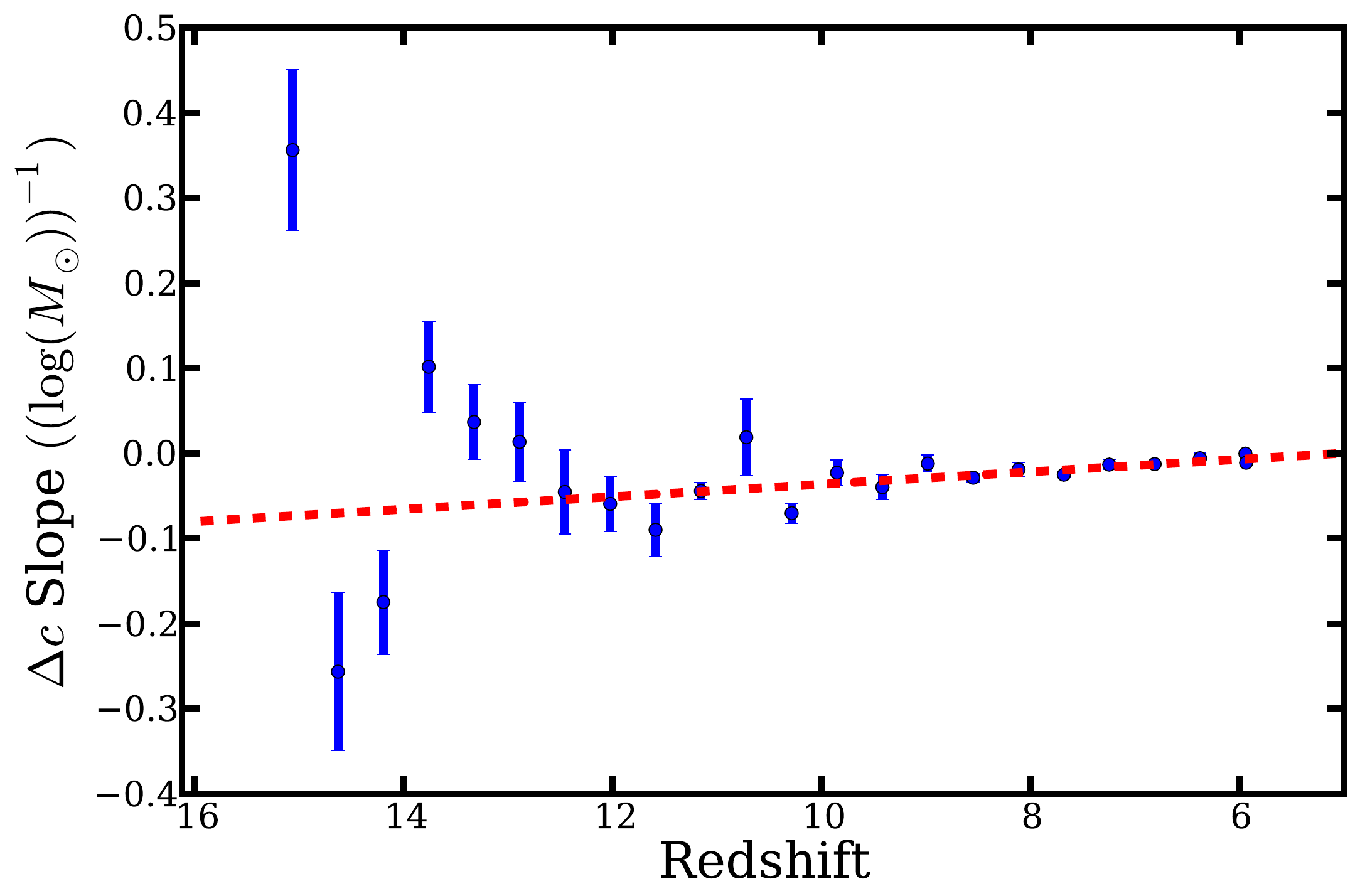}
	\end{subfigure}
	\caption[Slopes of the $\Delta q$ vs.\ $M_{\mathrm{vir,avg}}$ fit functions.]{\footnotesize Slopes of the $\Delta q$ vs.\ $M_{\mathrm{vir,avg}}$ fit functions.  The left and right panels correspond to the $\Delta M_{\mathrm{vir}}$ and $\Delta c$ plots in the left and right columns, respectively, of Figure~\ref{fig:delta-v-Mavg}.  Linear least-squares fits to the data are overplotted as red dashed lines.  Overall, we find a trend of positive and increasing slope with redshift for $\Delta M_{\mathrm{vir}}$ and negative and decreasing slope with redshift for $\Delta c$.  We find fit equations of $\mathrm{Slope} = (9.4 \pm 2.4) \times 10^{-4} z - (1.8 \pm 1.8) \times 10^{-3}$ for $\Delta M_{\mathrm{vir}}$ and $\mathrm{Slope} = -(7.3 \pm 1.9) \times 10^{-3} z + (3.7 \pm 1.4) \times 10^{-2}$ for $\Delta c$.  Snapshots at very high redshift, $z \gtrsim 14$ for $\Delta M_{\mathrm{vir}}$ and $z \gtrsim 13$ for $\Delta c$, begin to deviate from these trends.  However, it is uncertain if this deviation is significant due to the low number statistics of our sample at such high $z$.}
	\label{fig:slopes_delta-v-Mavg}
\end{figure*}

We consider $\Delta M_{\mathrm{vir}}$ and $\Delta c$ as a function of average halo mass $M_{\mathrm{vir,avg}} = (M_{\mathrm{vir},\lpt} + M_{\mathrm{vir},\za}) / 2$ for three representative timesteps in Figure~\ref{fig:delta-v-Mavg}.  The data are binned in average virial mass, for which means and standard deviations are provided as the black points and black dotted curves, respectively.  The error bars on the black points represent the uncertainty in the mean and are the standard deviation divided by the number of halos in that bin.  We additionally bin the data in rectangular bins on a 2-D grid with a logarithmic color map to feature the entire distribution of the data.  Linear fits to the bin means are overplotted in magenta.

We find that $\Delta M_{\mathrm{vir}}$ tends to increase with increasing $M_{\mathrm{vir,avg}}$ for most snapshots.  \lpt\ halos are consistently more massive than their \za\ counterparts, and, aside from the highest redshift snapshots, this difference increases with average halo mass.  While less massive halo pairs have a larger spread in the difference in \lpt\ and \za\ mass, more massive halo pairs are consistently heavier in \lpt\ than in \za.  At redshift 14.7, we find a transition between negative and positive slopes, and here the fit is $\Delta M_{\mathrm{vir}} = -(0.5 \pm 1.5) \times 10^{-2} \log(M_{\mathrm{vir,avg}}) + (0.15 \pm 0.12)$.   The slope of the fit lines then become positive and trends back towards zero as we progress in redshift, with a fit of $\Delta M_{\mathrm{vir}} = (3.49 \pm 0.99) \times 10^{-3} \log(M_{\mathrm{vir,avg}}) - (6.8 \pm 8.3) \times 10^{-3}$ by $z = 6$.

We additionally find a trend for more massive halo pairs to be more concentrated in \za.  This trend is somewhat stronger than for $\Delta M_{\mathrm{vir}}$, but again, high $z$ snapshots differ from the trend.  The fit equations for $z = 15$ and $z = 6$ are $\Delta c = -(0.256 \pm 0.093) \log(M_{\mathrm{vir,avg}}) + (2.07 \pm 0.76)$ and $\Delta c = -(1.10 \pm 0.31) \times 10^{-2} \log(M_{\mathrm{vir,avg}}) + (0.103 \pm 0.026)$, respectively.  The negative slope for most of the redshift range might be expected, as halo concentration is expected to decrease with increasing mass for all but the largest halos, where the concentration begins to increase with increasing mass \citep{2011ApJ...740..102K, 2012MNRAS.423.3018P}, and we find that $\Delta M_{\mathrm{vir}}$ increases with average mass for all but the highest redshift snapshots.  The turnover in halo concentrations displayed in \citet{2011ApJ...740..102K} and \citet{2012MNRAS.423.3018P} should be relatively inconsequential for our simulations, as we have a significantly smaller box size, and thus a smaller maximum halo mass.  Additionally, our most massive halos account for a very small percentage of the total halo population, causing the larger number of small halos to be more significant in the resulting fits.  The data have a larger variance than $\Delta M_{\mathrm{vir}}$ by a factor of $\sim 2$.  Again, mass dependence is smallest by $z = 6$.  To reconcile these trends with the symmetrical concentration distributions of Figure~\ref{fig:diff-hist}, we note that the trends in mass may be obscured by integration across the entire mass range and still result in overall $\Delta c$ distributions symmetric about zero.  Additionally, the histograms of Figure~\ref{fig:diff-hist} may be swamped by the large number of low mass halos, which masks the large difference in concentration seen here.

\begin{table}[t]
	\centering
	\caption{Coefficients for linear least squares fits from Figure~\ref{fig:slopes_delta-v-Mavg}.}
	\begin{tabular}{ c  r  r }
		\toprule
		                           &  \multicolumn{1}{c}{$A$}            &  \multicolumn{1}{c}{$B$} \\
		\cmidrule(l){2-3}
		$\Delta M_{\mathrm{vir}}$  &   $(9.4 \pm 2.4) \times 10^{-4}$  &  $(-1.8 \pm 1.8) \times 10^{-3}$ \\
		$\Delta c$                 &  $(-7.3 \pm 1.9) \times 10^{-3}$  &   $(3.7 \pm 1.4) \times 10^{-2}$ \\
		\bottomrule
	\end{tabular}
	\label{tab:coeffs_slopes}
\end{table}

The slopes of the fits to the $\Delta q$ vs.$M_{\mathrm{vir,avg}}$ data are plotted in Figure~\ref{fig:slopes_delta-v-Mavg}.  Linear least-squares fits are overplotted as red dashed lines.  We find a trend for there to be more $\Delta q$ dependence on $M_{\mathrm{vir,avg}}$ with increasing redshift, except for the highest $z$ snapshots, where the trends seem to reverse.  Coefficients $A$ and $B$ for the fit equation $\mathrm{Slope} = A z + B$ are listed in Table~\ref{tab:coeffs_slopes}.  The data are well-fit by the best fit line for most of the redshift range, except for $z \gtrsim 14$ for $\Delta M_{\mathrm{vir}}$ and $z \gtrsim 13$ for $\Delta c$, which begin to deviate from the trend.  While this may simply be due to the fluctuations inherent when dealing with the low number of matched halos available in our sample at these very high redshifts, a shift to positive slope for concentration may be expected.  At these redshifts, only the most massive halos halos fall above our particle threshold, whereas at later redshift, the large number of small halos can overwhelm the statistics.  These massive halos are most affected by high redshift differences due to initialization and may retain larger \lpt\ concentrations due to earlier formation.

%~~~~~~~~~~~~~~~~~~~~~~~~~~~~~~~~~~~~~~~~~~~~~~~~~~~~~~~~~~~~~~~~~~~~~~~~~~~~~~~
\subsection{A census of halo population differences}
%~~~~~~~~~~~~~~~~~~~~~~~~~~~~~~~~~~~~~~~~~~~~~~~~~~~~~~~~~~~~~~~~~~~~~~~~~~~~~~~

\begin{figure*}[t]
	\centering
	\begin{subfigure}{}
		\includegraphics[width=0.48\linewidth]{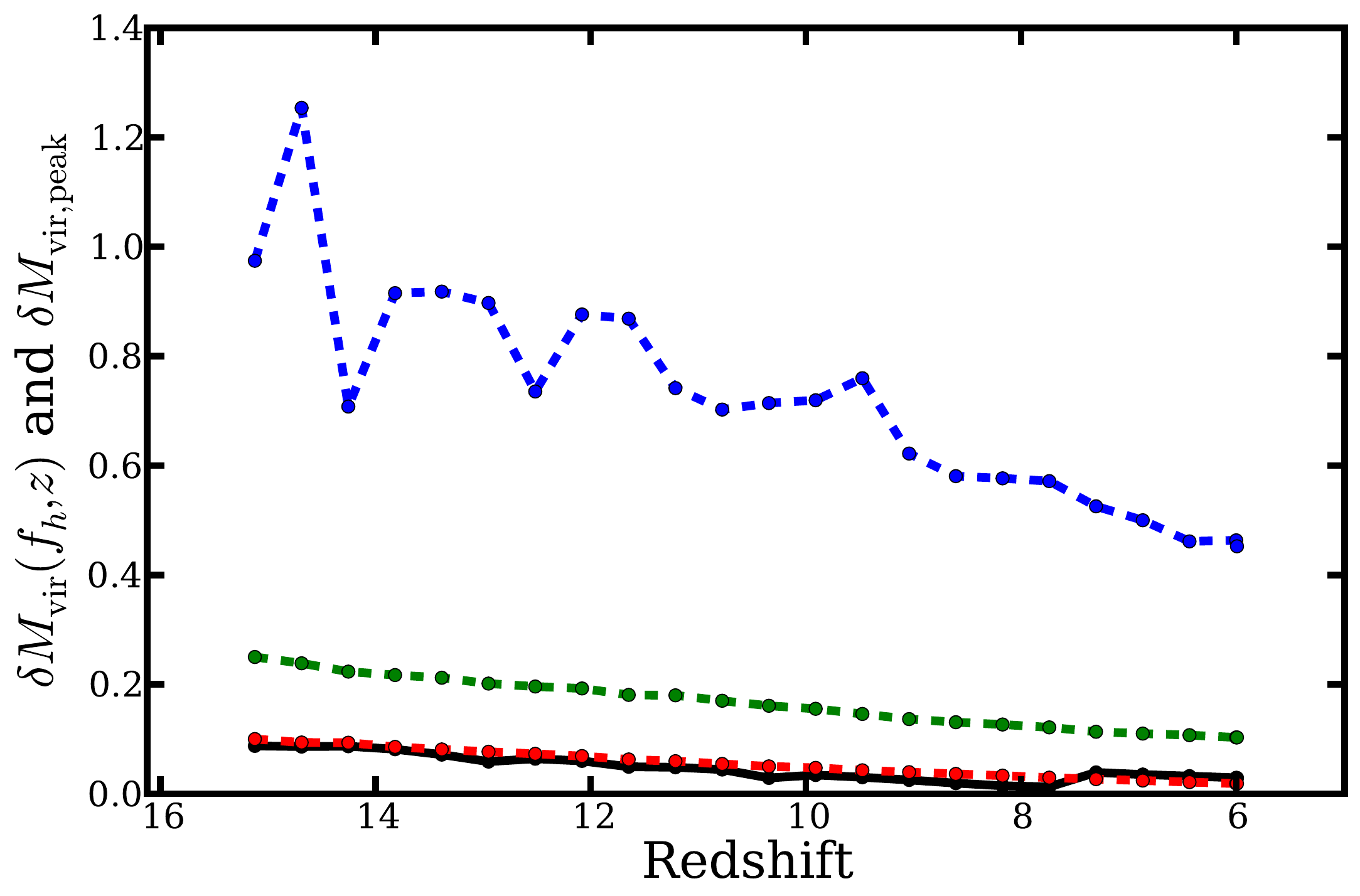}
	\end{subfigure}
	~
	\begin{subfigure}{}
		\includegraphics[width=0.48\linewidth]{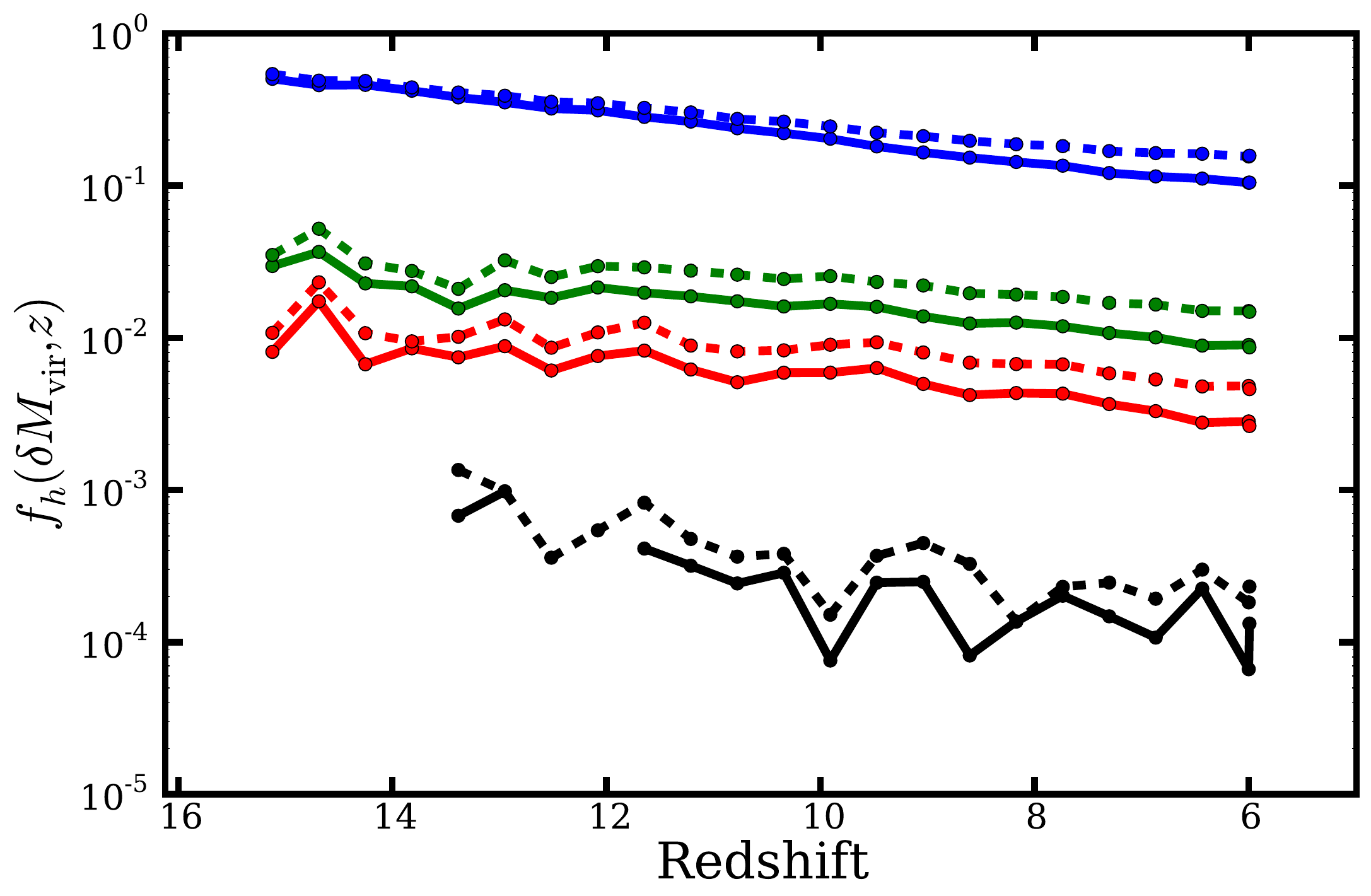}
	\end{subfigure}
	\\
	\begin{subfigure}{}
		\includegraphics[width=0.48\linewidth]{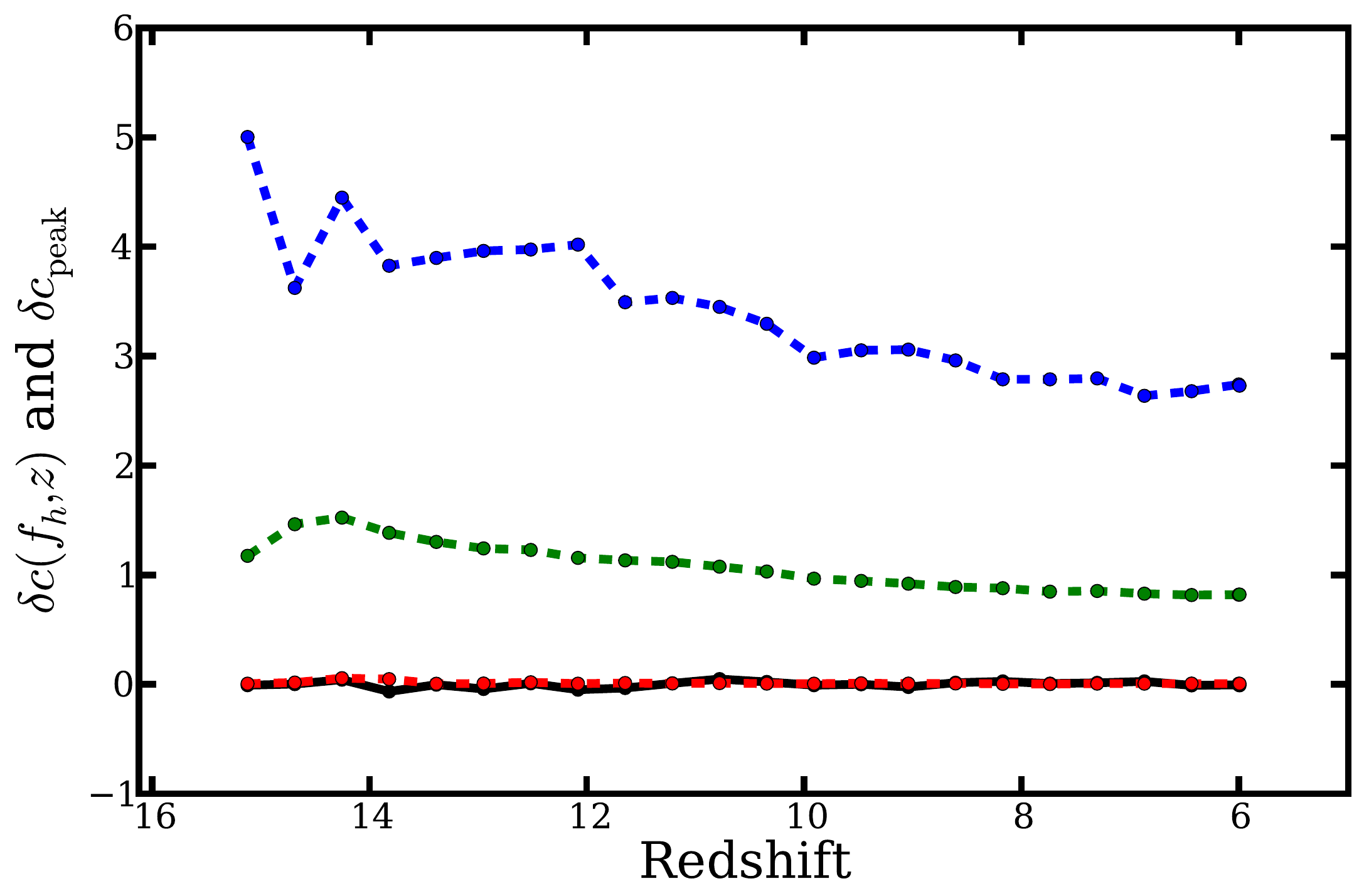}
	\end{subfigure}
	~
	\begin{subfigure}{}
		\includegraphics[width=0.48\linewidth]{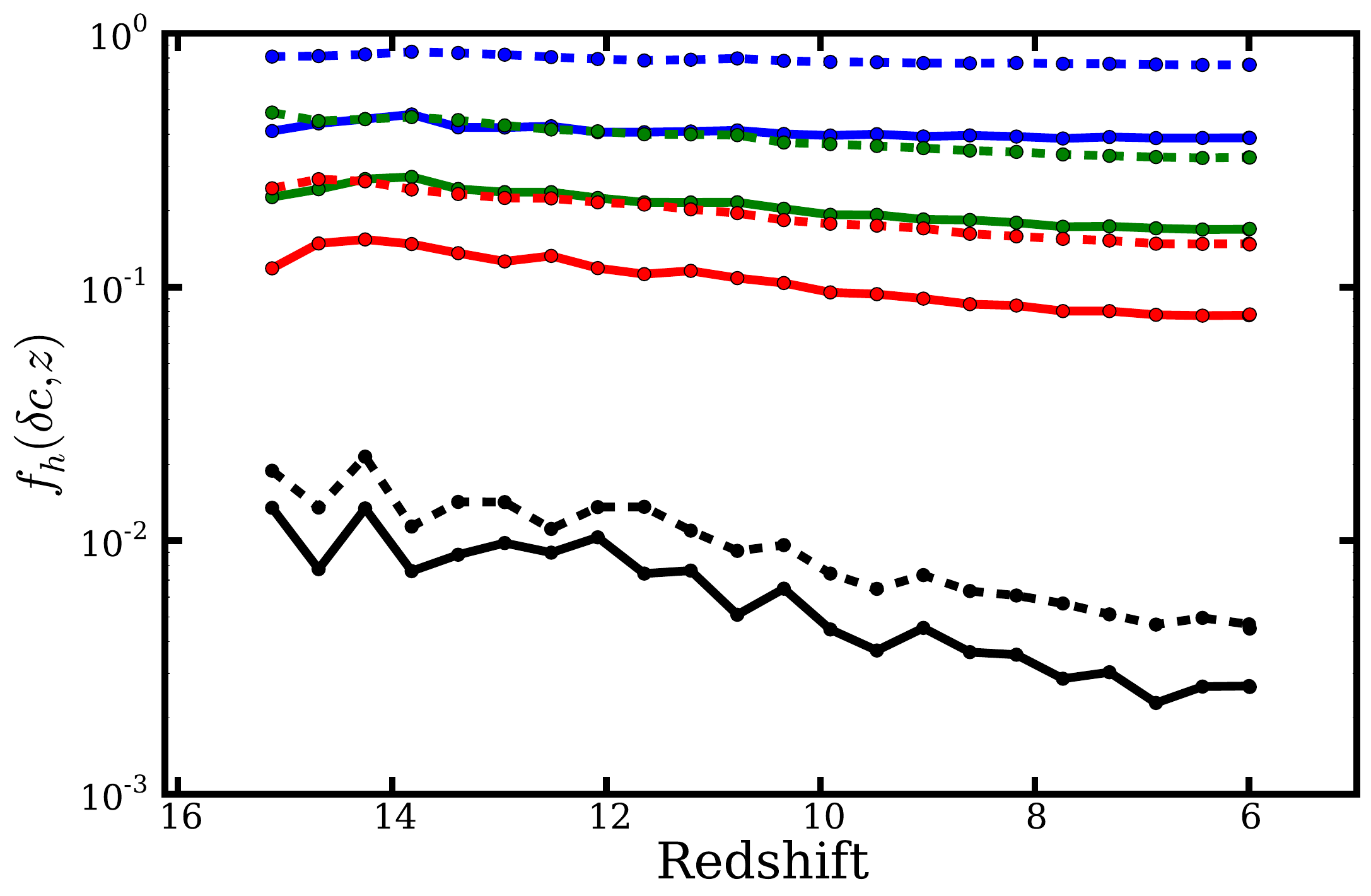}
	\end{subfigure}
	\caption[Statistics for distributions of $\delta q$ as functions of redshift]{\footnotesize Statistics for distributions of $\delta M_{\mathrm{vir}}$ (\textit{top row}) and $\delta c$ (\textit{bottom row}) as functions of redshift.  \textit{Left column}:  The $\delta q$ of the peak of the distribution (black curve), and the $\delta q$ where 50\% (red dashed curve), 10\% (green dashed curve), and 1\% (blue dashed curve) of the halos fall at or above $\delta q$.  As with distributions of $\Delta M_{\mathrm{vir}}$, $\delta M_{\mathrm{vir}}$ has the largest positive displacement at high redshift and steadily decreases throughout the simulation.  Additionally, $\delta c$ maintains a peak near zero and has a spread much larger than that of $\delta M_{\mathrm{vir}}$.  \textit{Right column}:  The fraction of halos with $\delta q$ greater than 0.10 (solid blue curve), 0.50 (solid green curve), 1.00 (solid red curve), and 4.00 (solid black curve).  The dashed curves additionally count halo pairs with $\delta q$ lower than the corresponding equivalent displacements of -0.09, -0.33, -0.50, and -0.80, respectively (see Equation~\ref{eq:equivalent_q_prime}).  We find that 50\% of \lpt\ halos are at least 10\% more massive than their \za\ companions at $z = 15$, reducing to 10\% by $z = 6$.  Halos in \lpt\ are at least twice as concentrated for 12\% of halos at $z = 15$ and 7.8\% of halos at $z = 6$.}
	\label{fig:frac_err_trends}
\end{figure*}

As our distributions of $\Delta q$ rely on the average quantity $q_{\mathrm{avg}} = (q_{\lpt} + q_{\za}) / 2$ for normalization, it can be difficult to extract certain statistics, such as the fraction of halo pairs differing by a certain amount between \lpt\ and \za\ simulations.  To address this, for this section, we redefine our difference distributions to instead use $q_{\za}$ as the normalization factor (see Equation~\ref{eq:delta_prime_q}).  In Figure~\ref{fig:frac_err_trends}, we plot, as functions of redshift, statistics derived from these alternate fractional difference distributions $\delta M_{\mathrm{vir}}$ and $\delta c$.  In the left column, we plot the $\delta q$ of the peak of the distribution along with the $\delta q$ where various percentages of the halo pairs fall at or above $\delta q$.

As the $\delta q$ value of the peak of the distribution is the location of the mode, it represents the most typical halo pair.  While concentration differences remain close to zero throughout the simulation, the mass difference peak moves from a $\delta M_{\mathrm{vir}}$ of $9 \times 10^{-2}$ at $z = 15$ to $3 \times 10^{-2}$ at $z = 6$.  The 1\% of halo pairs with the largest excess \lpt\ mass have \lpt\ mass at least twice \za\ mass at $z = 15$ and 1.5 times \za\ mass at $z = 6$.  For concentration, the 1\% most \lpt\ concentrated halo pairs differ by at least a factor of 6 at $z = 15$ and 4 at $z = 6$.

In the right column of Figure~\ref{fig:frac_err_trends}, we plot the fraction of halos $f_{h}$ that fall outside various $\delta q$ values.  The solid curves represent halo pairs that have $\delta q$ greater than or equal to the listed values, i.e., the fraction of halo pairs where the \lpt\ halo has a virial mass or concentration that is at least 1.1, 1.5, 2.0, or 5.0 times that of its corresponding \za\ halo.  The dashed curves represent the fraction of halo pairs where one halo has a virial mass or concentration at least 1.1, 1.5, 2.0, or 5.0 times that of its companion, regardless of whether the \lpt\ or \za\ value is higher.

We find that half of halo pairs are at least 10\% more massive in \lpt\ at $z = 15$.  By $z = 6$, this has fallen to 10\%.  Furthermore, 1\% are at least twice as massive in \lpt\ at $z = 15$, and by $z = 6$, this has only reduced to 0.3\%.  Halos in \lpt\ are at least twice as concentrated as their \za\ counterparts for at least 12\% of the halo population at $z = 15$ and at least 8\% by $z = 6$.  Halo pairs that are at least 5 times as concentrated in \lpt\ make up 1.3\% of the sample at $z = 15$ and 0.3\% at $z = 6$.

If we consider only the difference in properties between paired halos, regardless of whether the \lpt\ or \za\ halo has the higher mass or concentration, we include an even larger percentage of the population.   We find 54\% of the halo pairs differ in mass by at least 10\% at $z = 15$, with 16\% differing by $z = 6$.  Halos that are at least twice as massive in either \lpt\ or \za\ account for 1.1\% at $z = 15$ and 0.5\% at $z = 6$.  Halos that are at least twice as concentrated in either \lpt\ or \za\ account for 25\% at $z = 15$ and 15\% at $z = 6$.

%%%%%%%%%%%%%%%%%%%%%%%%%%%%%%%%%%%%%%%%%%%%%%%%%%%%%%%%%%%%%%%%%%%%%%%%%%%%%%%%
%
% Discussion
%
%%%%%%%%%%%%%%%%%%%%%%%%%%%%%%%%%%%%%%%%%%%%%%%%%%%%%%%%%%%%%%%%%%%%%%%%%%%%%%%%

\section{Discussion}
\label{sec:discussion}

%%%%%%%%%%%%%%%%%%%%%%%%%%%%%%%%%%%%%%%%%%%%%%%%%%%%%%%%%%%%%%%%%%%%%%%%%%%%%%%%

%~~~~~~~~~~~~~~~~~~~~~~~~~~~~~~~~~~~~~~~~~~~~~~~~~~~~~~~~~~~~~~~~~~~~~~~~~~~~~~~
% Recapitulation
%~~~~~~~~~~~~~~~~~~~~~~~~~~~~~~~~~~~~~~~~~~~~~~~~~~~~~~~~~~~~~~~~~~~~~~~~~~~~~~~

As we evolve our DM halo population from our initial redshift to $z = 6$, we find that simulation initialization with \lpt\ can have a significant effect on the halo population compared to initialization with \za.  The second order displacement boost of \lpt\ provides a head start on the initial collapse and formation of DM halos.  This head start manifests itself further along in a halo's evolution as more rapid growth and earlier mergers.  \lpt\ halos are, on average, more massive than their \za\ counterparts at a given redshift, with a maximum mean $\Delta M_{\mathrm{vir}}$ of $(9.3 \pm 1.2) \times 10^{-2}$ at $z = 15$.  The larger mass for \lpt\ halos is more pronounced for higher mass pairs, while \lpt\ halo concentration is larger on the small mass end.  Both mass and concentration differences trend towards symmetry about zero as halos evolve in time, with the smallest difference observed at the end of the simulations at $z = 6$, with a mean $\Delta M_{\mathrm{vir}}$ of $(1.79 \pm 0.31) \times 10^{-2}$.  Casual extrapolation of our observed trends with redshift to today would indicate that, barring structure like massive clusters that form at high redshift, \lpt\ and \za\ would produce very similar halo populations by $z = 0$.  However, the larger differences at high redshift should not be ignored.

%~~~~~~~~~~~~~~~~~~~~~~~~~~~~~~~~~~~~~~~~~~~~~~~~~~~~~~~~~~~~~~~~~~~~~~~~~~~~~~~
% Implications
%~~~~~~~~~~~~~~~~~~~~~~~~~~~~~~~~~~~~~~~~~~~~~~~~~~~~~~~~~~~~~~~~~~~~~~~~~~~~~~~

The earlier formation times and larger masses of halos seen in \lpt-initialized simulations could have significant implications with respect to early halo life during the Dark Ages.  Earlier forming, larger halos affect the formation of Pop-III stars, and cause SMBHs to grow more rapidly during their infancy \citep{2012ApJ...761L...8H} and produce more powerful early AGN.  The epoch of peak star formation may also be shifted earlier.  This could additionally increase the contribution of SMBHs and early star populations to the re-ionization of the universe.  Larger early halos may also increase clustering, speed up large scale structure formation, and influence studies of the high-$z$ halo mass function, abundance matching, gas dynamics, and galaxy formation.

In these discussions, it is important to note that it is wrong to assume that the \za\ halo properties are the ``correct'' halo properties, even in a statistical sense.  While halo mass suggests the most obvious shortcoming of \za\ simulations, even properties such as concentration---that show little difference on average between \lpt\ and \za---can have large discrepancies on an individual halo basis.  Failure to consider uncertainties in halo properties for high $z$ halos in \za\ simulations can lead to catastrophic errors.

%~~~~~~~~~~~~~~~~~~~~~~~~~~~~~~~~~~~~~~~~~~~~~~~~~~~~~~~~~~~~~~~~~~~~~~~~~~~~~~~
% Caveats
%~~~~~~~~~~~~~~~~~~~~~~~~~~~~~~~~~~~~~~~~~~~~~~~~~~~~~~~~~~~~~~~~~~~~~~~~~~~~~~~

We note a few caveats with our simulations and analysis.  We did not exclude substructure when determining the properties of a halo, and although this would not change the broad conclusions herein, care must be taken when comparing to works which remove subhalo particles in determining halo mass and concentration.  Halo matching is not perfect, as it is based on one snapshot at a time, and may miss-count halos due to merger activity and differences in merger epochs.  However, we believe this effect to be minor.  While we compared \rockstar's output with our own fitting routines and found them to broadly agree, \rockstar\ does not provide goodness of fit parameters for its NFW profile fitting and $R_{\mathrm{s}}$ measurements.  It also may be debated whether it makes sense to even consider concentration of halos at high redshift which are not necessarily fully virialized.

As \rockstar\ does not provide goodness-of-fit parameters for its internal density profile measurements used to derive concentration, error estimates for concentration values of individual halos are unknown.  Additionally, proper density profile fitting is non-trivial, as the non-linear interactions of numerical simulations rarely result in simple spherical halos that can be well described using spherical bins.  Halo centering issues may also come into play, although \rockstar\ does claim to perform well in this regard.

We use a simulation box size of only (10 Mpc)$^{3}$.  This is too small to effectively capture very large outlier density peaks.  We would, however, expect these large uncaptured peaks to be most affected by \lpt\ initialization, so the effects presented here may even be dramatically underestimated.  Additionally, a larger particle number would allow us to consider smaller mass halos than we were able to here, and to better resolve all existing structure.  A higher starting redshift could probe the regime where \lpt\ initialization contributes the most.  It would also be of interest to evolve our halo population all the way to $z = 0$.  The addition of baryons in a fully hydrodynamical simulation could also affect halo properties.  These points may be addressed in future studies.

%%%%%%%%%%%%%%%%%%%%%%%%%%%%%%%%%%%%%%%%%%%%%%%%%%%%%%%%%%%%%%%%%%%%%%%%%%%%%%%%
%
% Conclusion
%
%%%%%%%%%%%%%%%%%%%%%%%%%%%%%%%%%%%%%%%%%%%%%%%%%%%%%%%%%%%%%%%%%%%%%%%%%%%%%%%%

\section{Conclusion}
\label{sec:conclusion}

%%%%%%%%%%%%%%%%%%%%%%%%%%%%%%%%%%%%%%%%%%%%%%%%%%%%%%%%%%%%%%%%%%%%%%%%%%%%%%%%

We analyzed three \lpt\ and \za\ simulation pairs and tracked the spherical overdensity dark matter halos therein with the 6-D phase space halo finder code \rockstar\ to compare the effect of initialization technique on properties of particle--matched dark matter halos from $z = 300$ to $z = 6$.  This approach allowed us to directly compare matching halos between simulations and isolate the effect of using \lpt\ over \za.  In summary, we found the following:

\begin{itemize}

	\item  \lpt\ halos get a head start in the formation process and grow faster than their \za\ counterparts.  Companion halos in \lpt\ and \za\ simulations may have offset merger epochs and differing nuclear morphologies.

	\item  \lpt\ halos are, on average, more massive than \za\ halos.  At $z = 15$, the mean of the $\Delta M_{\mathrm{vir}}$ distribution is $(9.3 \pm 1.2) \times 10^{-2}$, and 50\% of \lpt\ halos are at least 10\% more massive than their \za\ companions.  By $z = 6$, the mean $\Delta M_{\mathrm{vir}}$ is $(1.79 \pm 0.31) \times 10^{-2}$, and 10\% of \lpt\ halos are at least 10\% more massive.

	\item  This preference for more massive \lpt\ halos is dependent on redshift, with the effect most pronounced at high $z$.  This trend is best fit by $\Delta M_{\mathrm{vir}} = (7.88 \pm 0.17) \times 10^{-3} z - (3.07 \pm 0.14) \times 10^{-2}$.
		
	\item  Earlier collapse of the largest initial density peaks causes the tendency for more massive \lpt\ halos to be most pronounced for the most massive halos, a trend that increases with redshift. We find a trend of $\Delta M_{\mathrm{vir}} = (1.03 \pm 0.46) \times 10^{-2} \log(M_{\mathrm{vir,avg}}) - (2.6 \pm 3.8) \times 10^{-2}$ for $z = 10$.  By $z = 6$, this has flattened to $\Delta M_{\mathrm{vir}} = (3.49 \pm 0.99) \times 10^{-3} \log(M_{\mathrm{vir,avg}}) - (6.8 \pm 8.3) \times 10^{-3}$.  As a function of redshift, the slopes of these equations are fit by $\mathrm{Slope} = (9.4 \pm 2.4) \times 10^{-4} z - (1.8 \pm 1.8) \times 10^{-3}$.

	\item  Halo concentration, on average, is similar for \lpt\ and \za\ halos.  However, even by the end of the dark ages, the width of the $\Delta c$ distribution---$\sigma_{\Delta c} = 0.551 \pm 0.026$ at $z = 6$---is large and indicative of a significant percentage of halos with drastically mismatched concentrations, despite the symmetrical distribution of $\Delta c$.  At $z = 15$, 25\% of halo pairs have at least a factor of 2 concentration difference, with this falling to 15\% by $z = 6$.

	\item  There is a trend for \za\ halos to be more concentrated than \lpt\ halos at high mass.  However, this trend seems to reverse above $z \sim 12$.  We find $\Delta c = -(0.256 \pm 0.093) \log(M_{\mathrm{vir,avg}}) + (2.07 \pm 0.76)$ at $z = 15$ and $\Delta c = -(1.10 \pm 0.31) \times 10^{-2} \log(M_{\mathrm{vir,avg}}) - (0.103 \pm 0.026)$ at $z = 6$.  The slopes of these equations, as a function of redshift, are fit by $\mathrm{Slope} = -(7.3 \pm 1.9) \times 10^{-3} z + (3.7 \pm 1.4) \times 10^{-2}$.  This is not visible in the symmetrical $\Delta c$ distributions, as the trends are roughly centered about zero and are washed away when integrated across the entire mass range.

\end{itemize}

We have found that choice of initialization technique can play a significant role in the properties of halo populations during the pre-reionization dark ages.  The early halo growth displayed \lpt\ simulations, or conversely the delayed halo growth arising from the approximations made in \za-initialized simulations, makes careful attention to simulation initialization imperative, especially for studies of halos at high redshift.  It is recommended that future \nbody\ simulations be initialized with \lpt, and that previous high-$z$ or high-mass halo studies involving \za-initialized simulations be viewed with the potential offsets in halo mass and concentration in mind.

This work was conducted using the resources of the Advanced Computing Center for Research and Education (ACCRE) at Vanderbilt University, Nashville, TN.  We also acknowledge the support of the NSF CAREER award AST-0847696.  We would like to thank the referee for helpful comments, as well as the first author's graduate committee, who provided guidance throughout this work.

\end{document}